%% file: main.tex
\documentclass[11pt]{article}
\usepackage{authblk}

\usepackage[utf8]{inputenc}
\usepackage[T1]{fontenc}
\usepackage{amsmath}
\usepackage{amsfonts}
\usepackage{amssymb}

\usepackage[normalem]{ulem} 

\usepackage{amsmath,xcolor,ulem}
\usepackage{amsfonts}
\usepackage{amssymb,todonotes}
\usepackage{amsthm}
\usepackage{graphicx}
\usepackage{marginnote}
\usepackage{todonotes,tikz,circuitikz}
\usetikzlibrary{angles,calc,quotes}
\usetikzlibrary {arrows.meta,bending,positioning}

\usepackage[colorlinks=true, allcolors=blue]{hyperref}

\usepackage[top=2.8cm,bottom=2.8cm,left=2.5cm,right=2.5cm]{geometry}
\usepackage{float}
\usepackage[algoruled]{algorithm2e}
\usepackage[font=footnotesize,width=\linewidth]{caption}

\newtheorem*{remark}{Remark}

\title{Data-driven identification of \\port-Hamiltonian DAE systems by Gaussian processes}

\author{Peter Zaspel\footnote{Research Group Scientific Computing and High Performance Computing, \href{mailto:zaspel@uni-wuppertal.de}{zaspel@uni-wuppertal.de}} \phantom{,} and Michael G\"unther\footnote{Research Group Applied and Computational Mathematics, \href{mailto:guenther@uni-wuppertal.de}{guenther@uni-wuppertal.de}}}
\affil{IMACM, School of Mathematics and Natural Sciences, \\ University of Wuppertal, Germany}

\begin{document}
\maketitle

\begin{tikzpicture}[remember picture,overlay]
	\node[anchor=north east,inner sep=20pt] at (current page.north east)
	{\includegraphics[scale=0.2]{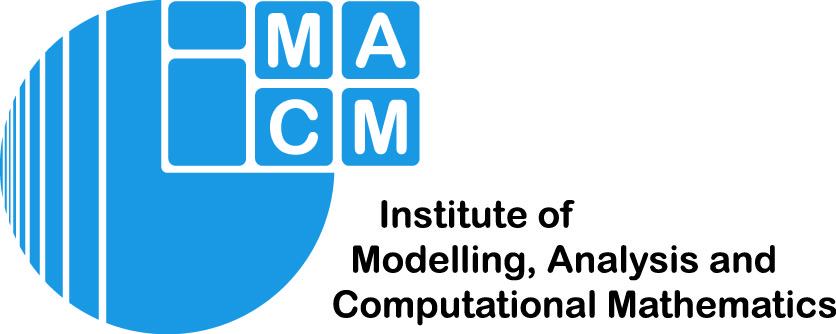}};
\end{tikzpicture}

\begin{abstract}
Port-Hamiltonian systems (pHS) allow for a structure-preserving modeling of dynamical systems. Coupling pHS via linear relations between input and output defines an overall pHS, which is structure preserving. However, in multiphysics applications, some subsystems do not allow for a physical pHS description, as (a) this is not available or (b) too expensive. Here, data-driven approaches can be used to deliver a pHS for such subsystems, which can then be coupled to the other subsystems in a structure-preserving way. In this work, we derive a data-driven identification approach for port-Hamiltonian differential algebraic equation (DAE) systems. The approach uses input and state space data to estimate nonlinear effort functions of pH-DAEs. As underlying technique, we us (multi-task) Gaussian processes. This work thereby extends over the current state of the art, in which only port-Hamiltonian ordinary differential equation systems could be identified via Gaussian processes. We apply this approach successfully to two applications from network design and constrained multibody system dynamics, based on pH-DAE system of index one and three, respectively. 
\end{abstract}

\begin{minipage}{0.9\linewidth}
 \footnotesize
\medskip

\end{minipage}

\section{Introduction}

In multiphysics modelling, one is faced with solving coupled systems of differential-algebraic equations (DAE), which describe the individual PDE subsystems after a semidiscretization with respect to space. The automatic generation of the subsystem models, usually based on a network approach, involves redundant information, and hence leads to DAE models instead of ODE models.
Structure preservation is mandatory here: if the single subsystems are dissipative or energy-preserving, for example, the joint system has to inherit these properties. Hence one has to choose a coupling approach which preserves these structures. 
Here, the port-Hamiltonian framework provides the right tool. It allows 
 for designing overall port-Hamiltonian systems (pHS) that inherit the structrual properties of the subsystems provided that  all subsystems are pHS and a   linear (power-conserving) interconnection between the inputs and outputs of the subsystems is provided~\cite{
    GuZa_vanDerSchaft06,
    mehrmannmorandin2019}.

 In realistic applications this approach reaches its limits if for a specific subsystem, either (a) no knowledge that would allow the definition of a physics-based pHS is available, or (b) one is forced to use user-specified simulation packages with no information of the intrinsic dynamics (which may follow a pHS dynamics or not), and thus only the input-output characteristics are available, or (c) a pHS description is available, but too expensive for being used in the overall simulation. 
%
 However, there is an interest to construct a model that is guaranteed to follow pHS properties in order to be used as part of a bigger coupled pHS system. 
 
 Data-driven models allow to overcome these limitations. In case of
 an unknown physical pHS model, measurement data is collected and used to fit a pHS surrogate model, while in the case of
 too expensive model evaluations or black-box simulation packages, data is collected from simulation runs. This surrogate model guarantees structural pHS properties, is cheap to be evaluated and can hence be efficiently used in realistic multiphysics applications.

Data-driven identification of dynamic pHS models data has been a topic of research during the last years, with a focus on ODE systems, especially on linear systems. For the latter  a frequency approach is obvious, see, for example~\cite{GuZa_benner2020identification}, where the Loewner framework has been proposed. Other approaches include 
a parameterization approach based on using unconstrained optimization solvers during identification~\cite{GuZa_SCH} or infer frequency response data using time-domain input-output data~\cite{GuZa_ChGB22}. Dynamic mode decompositions are used in~\cite{GuZa_MoNiUn22} for identifying a linear pH-ODE model based 
on  input, state-space  and output time-domain data and a given quadratic Hamiltonian. Further, a two-step identification procedure has been proposed in~\cite{GuZa_ChMH19} based on deriving first a best-fit linear state-space model and then inferring a nearest linear pH-ODE model. As a one-step procedure, two iterative gradient-based and adjoint-based approaches  have been derived in~\cite{GuZa_2,GuZa_23}. 

For nonlinear pH-ODE systems, Koopman theory is used in~\cite{GuZa_JUNKER2022389} to 
shift the nonlinear problem to a linear one. Regarding pH-DAE systems, 
\cite{GuZa_diss_Medianu} discusses an approach 
based on discrete-time input-output data, however, assuming data steeming from symplectic integration schemes. 
Another rather general approach is given by physics-informed neural networks (PINNs), which combine data-based and physics-based approaches. Such approaches have been introduced in ~\cite{GuZa_weiqi2021,GuZa_moya2023} for  applications in stiff chemical reaction kinetics (nonlinear pH-ODEs) and power network (nonlinear pH-DAEs). 

Recently, Beckers et al.~\cite{GuZa_9992733} have proposed an alternative approach for nonlinear pH-ODE systems based on Gaussian processes (GPs). Data-driven function approximation based on GPs, also often referred to as Gaussian process regression (GPR), is a well known statistics oriented machine learning approach that can easily deal with noisy data. This makes it an ideal candidate for realistic data-driven applications. Prior to the work of Beckers et al.~on pH-ODE identification, GP based approaches for Hamiltonian ODE identification \cite{GuZa_10.1063/5.0048129} and dissipative ODE system identification \cite{GuZa_NEURIPS2022_82f05a10} have been developed, in which the Hamiltonian of the system is found from input and state space data. These works share that the scalar Hamiltonian is modeled as a Gaussian process and learned via its gradient. Dissipation and the full pHS structure is treated by linear transformations and shifts of the GP, respectively. Beckers et al.~\cite{GuZa_9992733} have in particular shown that their data-driven approximation of the pH-ODE system fulfills the pHS properties, hence is structure preserving. Thereby, the GP based data-driven approach can both deal with noisy data and is structure preserving, hence becomes a very appealing method for data-driven pHS identification.

In this work, we extend prior work on data-driven pH-ODEs using Gaussian processes towards the case of nonlinear pH-DAE systems, which -- to the best of our knowledge -- has not been done before. In difference to prior approaches, we identify the vector-valued effort function of the pH-DAE system, i.e.~no longer the (gradient of) the scalar Hamiltonian. To this end, we have to treat the case of vector-valued function approximation, which we achieve by multi-task Gaussian processes~\cite{GuZa_10.5555/2981562.2981582}. Moreover, we need to cover the algebraic constraints in the system. In our numerical results, in which we cover a basic test cases from electric network design and constrained multibidy system dynamics, we empirically observe excellent approximations up to the data's discretization accuracy with only very few training samples.

Overall, our proposed approach (a) fills the gap between nonlinear pH-ODE and pH-DAE systems, (b) constructs models which are preserving structure exactly and (c) does not demand synthetic data arising from structure-preserving discretization as it can deal with noisy data.

The paper is organized as follows. In Section~\ref{sec:identificationOfPhsDaeSystems}, we describe the identification task by introducing pH-ODEs and stating the identification problem. Section~\ref{sec:dataDrivenIdentificationViaGaussianProcesses} introduces the novel data-driven identification of the effort function in pH-DAEs. To this end (multi-task) Gaussian process regression its application in the pH-DAE case is discussed with some implementation details. In Section~\ref{sec:numericalResults}, we introduce the two model systems and give a detailed analysis of prediction/identification errors of the novel method on some exemplary test data. This is followed by conclusions.




\section{Identification of pH-DAE systems}
\label{sec:identificationOfPhsDaeSystems}
\subsection{Port-Hamiltonian differential algebraic equations}
We consider port-Hamiltonian differential-algebraic  (short-hand: pH-DAE) systems~\cite{mehrmannmorandin2019} of the special structure
\begin{align}
\label{eq:GuZaPHSgeneralDAE}
    E(\mathbf{x}) \cdot \dot{\mathbf{x}} = (J(\mathbf{x})-  R(\mathbf{x}))  \mathbf{z}(\mathbf{x}) + B(\mathbf{x}) \mathbf{u}, & \, &
    \mathbf{y} =  B(\mathbf{x})^\top  \mathbf{z}(\mathbf{x})
\end{align}
without feed-through, with time-dependent state $\mathbf{x}$, input $\mathbf{u}$, output $\mathbf{y}$ and effort function $\mathbf{z}(\mathbf{x})$.  The solution depenent matrices $J(\mathbf{x})$, $R(\mathbf{x})$ and $B(\mathbf{x})$ denote the skew-symmetric structure, the positiv-semidefinit dissipation and the port matrix. In DAEs, the solution-dependent flow matrix $E(\mathbf{x})$ is singular. Then, if the compatability condition $E(\mathbf{x})^\top  \mathbf{z}(\mathbf{x}) = \nabla H(\mathbf{x})$ holds, with $H$ the Hamiltonian of the system, which is at least continously differentiable, the pH-DAE fulfills the dissipativity inequality
\begin{equation}
    \label{eq.dissinequ}
    \frac{d}{dt} H(\mathbf{x}) =  \nabla H(\mathbf{x})^\top \dot{\mathbf{x}} = - \nabla H(\mathbf{x})^\top R(\mathbf{x}) H(\mathbf{x}) + \mathbf{y}^\top \mathbf{u} \le \mathbf{y}^\top \mathbf{u}.
\end{equation}
The main advantage of PHS modeling is that coupled pH-DAE systems define again a pH-DAE system fulfilling a dissipativity inequality~\cite{GuMa_gu23}: 
consider $s$ subsystems of pH-DAE systems
$$
\begin{aligned}
 E_i  \frac{d}{dt} \mathbf{x_i} 
    =& (J_i(\mathbf{x}) - R_i(\mathbf{x})) \mathbf{z_i}(\mathbf{x_i})   + B_i(\mathbf{x}) \mathbf{u_i},
&\mathbf{y_i} =    B_i(\mathbf{x})^\top  \mathbf{z_i}(\mathbf{x_i}),
 \end{aligned}
$$
$i=1,\ldots,s$ with $J_i=-J_i^\top$ and $R_i=R_i^\top \ge 0$.
The inputs $\mathbf{u_i}$, the outputs $\mathbf{y_i}$ and $B_i(\mathbf{x_i})$ are split into
$$
    \label{eq:split-input-output}
    \mathbf{u_i}= \begin{pmatrix}
                \mathbf{\hat u_i} \\
                 \mathbf{\bar u_i}
        \end{pmatrix},
        \quad
    \mathbf{y_i}= \begin{pmatrix}
                 \mathbf{\hat y_i} \\
                \mathbf{\bar y_i}
        \end{pmatrix}, \quad 
      B_i = \begin{pmatrix}
                \hat B_i &
                \bar B_i
            \end{pmatrix}   
$$
according to external and coupling quantities.
The subsystems are coupled via external inputs and outputs by %
\begin{align*}
    \begin{pmatrix}
    \mathbf{\hat u_1} \\ \vdots \\ \mathbf{\hat u_k}
    \end{pmatrix} 
    + \hat C 
    \begin{pmatrix}
     \mathbf{\hat y_1} \\ \vdots \\  \mathbf{\hat y_k}
    \end{pmatrix} = 0, \qquad \hat C = - \hat C^\top. 
\end{align*}
These $s$ systems can be condensed to one large pH-DAE system
\begin{subequations}\label{cond.ph.dae}
\begin{align}
E(\mathbf{x}) \frac{d}{dt}  \mathbf{x}  =   (\tilde J( \mathbf{x}) -R( \mathbf{x}))   \mathbf{z}( \mathbf{x})  +  \bar B \mathbf{\bar u}, & \, & 
\mathbf{\bar y}  =  \bar B^{\top}  \mathbf{z}( \mathbf{x}) 
\end{align}
\end{subequations}
with $\tilde J=J - \hat B \hat C \hat B^{\top}$, the Hamiltonian $H(\mathbf{x})$ given by $H(\mathbf{x}):=\sum_{i=1}^s H_i(\mathbf{x_i})$ and the condensed quantities 
\begin{gather*}
 \mathbf{v}^{\top} =  (\mathbf{v_1}^{\top},\ldots, \mathbf{v_s}^{\top})
    \quad  %
        \text{ for } \mathbf{v} \in \{\mathbf{x}, \, \mathbf{\bar u}, \mathbf{\bar y}, \, \mathbf{z}\},
       \\
  F=\mbox{diag}\,(F_1,\ldots, F_s)
      \quad %
        \text{ for } F \in \{E,\,Q,\,J,\,R,\,\hat{B},\, \bar{B}\}.
\end{gather*}
Equation~\eqref{cond.ph.dae}
defines now a pH DAE of type~\eqref{eq:GuZaPHSgeneralDAE}.


For many pH-DAE systems arising in engineering and science applications 
the system matrix $E(\mathbf{x})$ is given in block-diagonal form of the type
$E(x)=$diag$(D,0)$ 
with a constant regular matrix  $D$. If we split $\mathbf{x}:=(\mathbf{x_1},\mathbf{x_2})$ and $\mathbf{z}(\mathbf{x}):=(\mathbf{z_1}(\mathbf{x_1},\mathbf{x_2}),\mathbf{z_2}(\mathbf{x_1},\mathbf{x_2}))$ with respect to the dimensions of the block structure above,  the compatability condition now reads 
$$
\begin{bmatrix}
    \mathbf{z_1}(\mathbf{x_1},\mathbf{x_2}) \\ 0
\end{bmatrix} =
\begin{bmatrix}
    D^{-\top} \cdot \nabla_{\mathbf{x_1}} H(\mathbf{\mathbf{x_1},\mathbf{x_2}}) \\
    \nabla_{\mathbf{x_2}} H(\mathbf{\mathbf{x_1},\mathbf{x_2}}) 
\end{bmatrix},
$$
and thus $H(\mathbf{x_1},\mathbf{x_2})
=H(\mathbf{x_1}).$
The differential index is given by one, if 
$$
\frac{\partial}{\partial \mathbf{x_2}}
\left(
\begin{bmatrix}
    0 & I 
\end{bmatrix}
\cdot \begin{bmatrix}
    ({J}(\mathbf{x})-{R}(\mathbf{x})) {\mathbf{z}}(\mathbf{x}) + {B}(\mathbf{x}) \mathbf{u}
\end{bmatrix} \right)
$$
is regular in a neighborhood of the solution. 

%
\begin{remark}
    The skew-symmetric structure matrix $J$ is usually defined by the structure or topology of the system, resp., and thus constant in most cases. Hence we set $J$ to be constant in the following. In addition, for simplicity, also $R$ and $B$ are assumed to be constant. 
\end{remark}


\subsection{Identification problem}
The identification task discussed in this paper is given as follows: 
for given data
$$\{(\mathbf{u}(t_i), \mathbf{x}(t_i)\}_{i=1}^{N_T}, \quad t_i\in [0,T] \, \forall \, i=1, \ldots N_T$$
the nonlinear effort function $\mathbf{z}$ has to be identified. For this, one has first to derive derivative data $\{\dot{\mathbf{x}}(t_i) \}_{i=1}^{N_T}$ from the given data (see section~\ref{sec.derdata} for details). Then~\eqref{eq:GuZaPHSgeneralDAE} yields for all time points $t_i$ the identity
$$
\mbox{diag}(D,0) \cdot \dot{\mathbf{x}}(t_i) - B \mathbf{u}(t_i) =
(J-R) \mathbf{z}(\mathbf{x}(t_i)),
$$
which allows for reconstructing  $\{\mathbf{z}(t_i)\}_{i=1}^{N_T}
$ provided that  $J(\mathbf{x})-R(\mathbf{x})$ is regular.
%
However, if $J-R$ is singular, then the non-trivial part of $\mathbf{z}(\mathbf{x})$ lying 
in the kernel of $J-R$ cannot be identified unless additional information is available.

However, the regularity of $J-R$ is often demanded in engineering and science applications, when a unique solution of the operation point analyis is required: the system should have a unique constant solution for a constant input, i.e., if $\dot{\mathbf{x}}$ is set to zero. In this case system~\eqref{eq:GuZaPHSgeneralDAE} yields the linear equation
$$
(J-R) \mathbf{z}(\mathbf{x}) = - B \mathbf{u},
$$
which has a unique solution only if $J-R$ is regular.



\subsection{Construction of derivative data}

\label{sec.derdata}

It remains to obtain derivative data $\{\dot{\mathbf{x}}(t_i) \}_{i=1}^{N_T}$
from the given data $\{(\mathbf{u}(t_i), \mathbf{x}(t_i)\}_{i=1}^{N_T}$. This can be done either by numerical finite differences or first constructing an interpolation function based on Gaussian processes, cubic splines or any other interpolation procedure, and then analytically differentiate the interpolation function and evaluate the derivative at $t_i$. The error will remain small, if the data is sampled at a sufficiently high rate. However, large differences between time points, caused by a low sampling rate or inappropriate choice of training data, will yield large approximation errors for the obtained derivatives.


\section{Data-driven identification via Gaussian Processes}
\label{sec:dataDrivenIdentificationViaGaussianProcesses}
We model the effort function $\mathbf{z}(\mathbf{x})$ as a vector-valued Gaussian process. In the following, standard Gaussian processes, their use in function approximation / regression and the necessary extension to vector-valued functions by multi-task Gaussian processes (MT-GP) are reviewed. Based on this, the identification of the effort function in the pH-DAE setting is introduced.
\subsection{Gaussian process regression}
Let us assume for now that we are interested to find a function $z:\mathbb{R}^d\rightarrow \mathbb{R}$ from (training) data $\mathcal{T}=\{(\mathbf{x}_i,y_i)\}_{i=1}^{N_T}$. We follow the common notation in GP literature and introduce the set of inputs as $\mathbf{X}=\{\mathbf{x}_i\}_{i=1}^{N_T}$. This allows to write the vector of outcomes of function $z$ on inputs $\mathbf{X}$ as $z(\mathbf{X})=[z(\mathbf{x}_1) \cdots 
 z(\mathbf{x}_{N_T})]^\top$. Similarly, the data vector is ${Y}= [y_1 \cdots y_{N_T}]^\top$.
GP regression starts from putting a Gaussian process as prior distribution on the to be found function $z$, hence $$z\sim \mathcal{GP}(m,k)\,,$$ where $m:\mathbb{R}^d\rightarrow \mathbb{R}$ is the mean function and $k:\mathbb{R}^d\times \mathbb{R}^d \rightarrow \mathbb{R}$ is the covariance of the Gaussian process. The covariance function is also often called \textit{kernel}. It is common in GP regression to manually choose the positive definite kernel function, where a typical choice is the Gaussian kernel $k(\mathbf{x},\mathbf{x}^\prime) = \exp\left(-\frac{\|\mathbf{x}-\mathbf{x}^\prime\|^2}{2\phi^2}\right)$ with scaling parameter $\phi$, which will be used throughout this work. With the choice of $z$ following a GP, we obtain
$$z(\mathbf{X}) \sim \mathcal{N}(m(\mathbf{X}),k(\mathbf{X},\mathbf{}))\,,$$
hence the vector of outcomes of function $z$ follows a multivariate normal distribution with mean $m(\mathbf{X})$, which is the vector of outcomes of $m$ applied to the set $\mathbf{X}$, and covariance matrix $k(\mathbf{X},\mathbf{X}) = (k_{ij})_{i,j=1}^{N_T}$, with $k_{ij}=k(\mathbf{x}_i,\mathbf{x}_j)$. In GP regression, it is common to use a zero mean, hence $m \equiv 0$, giving $z(\mathbf{X}) \sim \mathcal{N}(\mathbf{0},k(\mathbf{X},\mathbf{X}))$.

We are now interested to make predictions. To this end, we consider $N_{eval}$ evaluation points $\mathbf{X}_\ast = \{\mathbf{x}_{\ast,1}, \ldots \mathbf{x}_{\ast,{N_{eval}}}\}$, giving $z(\mathbf{X}_\ast)=[z(\mathbf{x}_{\ast,1}) \cdots z(\mathbf{x}_{\ast,N_{eval}})]^\top$. In a first step we restrict ourselves to noise-free observations, i.e.~have the interpolatory condition
\begin{equation}\label{eq:interpolatoryCondition}y_i = z(\mathbf{x}_i), \quad \mbox{for}\ i=1,\ldots, N_T\,,
\end{equation}
thus $Y = z(\mathbf{X})$. Then, the joint distribution of the training outputs $Y$ and the predicted outputs ${Y}_\ast$ is (using zero mean)
$$\left[\begin{array}{c} {Y}\\{Y}_\ast
\end{array}\right] = \left[\begin{array}{c} z(\mathbf{X})\\{Y}_\ast
\end{array}\right]
 \sim \mathcal{N}\left(
\mathbf{0},
\left[\begin{array}{cc} k(\mathbf{X},\mathbf{X}) & k(\mathbf{X},\mathbf{X}_\ast)\\
k(\mathbf{X}_\ast,\mathbf{X}) & k(\mathbf{X}_\ast, \mathbf{X}_\ast)
\end{array}\right]
 \right)\,.
$$
Here, we have used, e.g., the notation $k(\mathbf{X},\mathbf{X}_{\ast})$ to describe the kernel matrix of size $N_T\times N_{eval}$ with entries $k_{ij}(\mathbf{x}_i,\mathbf{x}_{\ast,j})$. \textit{Conditioning} and a simple linear algebra argument then allows to derive the posterior distribution for the predicted outputs as
\begin{align*}
{Y}_\ast| z, \mathbf{X}_\ast,\mathbf{X}, \phi \sim \mathcal{N}( &k(\mathbf{X}_\ast,\mathbf{X}) k(\mathbf{X},\mathbf{X})^{-1} {Y}, \\
& k(\mathbf{X}_\ast,\mathbf{X}_\ast) - k(\mathbf{X}_\ast,\mathbf{X})  k(\mathbf{X},\mathbf{X})^{-1} k(\mathbf{X},\mathbf{X}_\ast))\,.
\end{align*}
The predictions are given via the mean of the posterior distribution, while the covariance gives a measure on the uncertainty in the prediction.

To capture and appropriatly regularize the case of noisy data, we further assume a Gaussian noise model on the observed data. In other words, instead of the interpolatory condition in \eqref{eq:interpolatoryCondition}, it is assumed to have 
$$y_i = z(\mathbf{x}_i) + \epsilon_i, \quad \mbox{with}\ \epsilon_i \sim\mathcal{N}(0,\sigma^2), \quad \mbox{for}\ i=1,\ldots,N_T\,.$$
Then the observed data $y_i$, conditioned to the GP $z$, the input $\mathbf{x}_i$ and the noise variance $\sigma^2$ follows a normal distribution with mean $z(\mathbf{x}_i)$ and variance $\sigma^2$, i.e.
$$y_i | z,\mathbf{x}_i,\sigma^2,\phi \sim \mathcal{N}(z(\mathbf{x}_i),\sigma^2)\,.$$
It is immediate that the vector of observed outputs ${Y}$ then has the corresponding likelihood 
\begin{equation}\label{eq:dataLikelihood}{Y} | z,\mathbf{X},\sigma^2,\phi \sim \mathcal{N}(z(\mathbf{X}),\sigma^2 I_{N_T})\,,
\end{equation}
with $I_{N_T}$ the $N_T$-dimensional identity matrix. It can be shown that then the joint distribution of the training outputs ${Y}$ and the predicted outputs ${Y}_\ast$ is 
$$\left[\begin{array}{c} {Y}\\{Y}_\ast
\end{array}\right]
 \sim \mathcal{N}\left(
\mathbf{0},
\left[\begin{array}{cc} k(\mathbf{X},\mathbf{X}) + \sigma^2 I_{N_T} & k(\mathbf{X},\mathbf{X}_\ast)\\
k(\mathbf{X}_\ast,\mathbf{X}) & k(\mathbf{X}_\ast, \mathbf{X}_\ast)
\end{array}\right]
 \right)\,,
$$
and conditioning gives 
\begin{align*}
{Y}_\ast| z, \mathbf{X}_\ast,\mathbf{X}, \sigma^2, \phi \sim \mathcal{N}( &k(\mathbf{X}_\ast,\mathbf{X}) [k(\mathbf{X},\mathbf{X})+ \sigma^2 I_{N_T}]^{-1} {Y}, \\
& k(\mathbf{X}_\ast,\mathbf{X}_\ast) - k(\mathbf{X}_\ast,\mathbf{X})  [k(\mathbf{X},\mathbf{X})+ \sigma^2 I_{N_T}]^{-1} k(\mathbf{X},\mathbf{X}_\ast))\,.
\end{align*}
The hyperparameter $\phi$ of the chosen kernel and the noise variance $\sigma^2$ is typically estimated jointly in the inference process. A common approach for this is to use a maximum likelihood estimator. That is, parameters $\sigma^2,\phi$ are found that maximize the \textit{marginal (log) likelihood} of the outputs $\mathbf{Y}$, i.e.~$\log p({Y}|\mathbf{X},\sigma^2,\phi)$. In difference to eq.~\eqref{eq:dataLikelihood}, the \textit{marginal} likelihood has been marignalized wrt.~the GP $z$, i.e.~$z$ is integrated out. As further discussed in \cite[Chapter~2]{Rasmussen2005}, the log marginal likelihood of $\mathbf{Y}$ is given by
$$\log p({Y}|\mathbf{X},\sigma^2,\phi) = -\frac{1}{2} {Y}^\top [k(\mathbf{X},\mathbf{X})+ \sigma^2 I_{N_T}]^{-1} {Y} - \frac{1}{2} \log | k(\mathbf{X},\mathbf{X})+ \sigma^2 I_{N_T} | - \frac{N_T}{2} \log 2\pi\,.$$
While GP regression for fixed hyperparameters has a computational complexity of $O(N^3_T)$ due to the inversion of the covariance or kernel matrix $k(\mathbf{X},\mathbf{X})$, the joint estimation of the hyperparameters by the maximum likelihood approach increases the computational complexity to $O(N_{opt} N_T^3)$, where $N_{opt}$ is the number of optimization steps required to maximize the (log) marginal likelihood.

\subsection{Multi-task Gaussian process regression}
To be able to identify the \textit{vector-valued} effort function $z:\mathbb{R}^d\rightarrow \mathbb{R}^D$, we require vector-valued Gaussian processes. Here, we start from data $\mathcal{T}=\{(\mathbf{x}_i,\mathbf{y}_i)\}_{i=1}^{N_T}$, with $\mathbf{y}_i=\left[y^{(1)}_i \cdots y_i^{(D)}\right]^\top\in\mathbb{R}^D$, for $i=1,\ldots, N_T$.
While  $\mathbf{X}=\{\mathbf{x}_i\}_{i=1}^{N_T}$, is as before, we introduce $\mathbf{z}(\mathbf{X})$ as the $N_T\cdot D$-dimensional vector of outcomes of the vector-valued $\mathbf{z}$ for all inputs. The entries of this vector are sorted such that first all first components, then all second components, etc.~are concatenated, i.e.
$$\mathbf{z}(\mathbf{X})=\left[z^{(1)}(\mathbf{x}_1) \cdots z^{(1)}(\mathbf{x}_{N_T}) z^{(2)}(\mathbf{x}_1) \cdots z^{(2)}(\mathbf{x}_{N_T}) \cdots\right]^\top\,.$$
The $N_T\cdot D$-dimensional data vector $\mathbf{Y}$ then has the same ordering, which is
$$\mathbf{Y}=\left[y^{(1)}_1 \cdots y^{(1)}_{N_T} y^{(2)}_1 \cdots y^{(2)}_{N_T} \cdots\right]^\top\,.$$
In the vector-valued case, we have
$$\mathbf{z}\sim \mathcal{GP}(\mathbf{m},\mathbf{k})\,,$$
where $\mathbf{m}:\mathbb{R}^d\rightarrow \mathbb{R}^D$ is the vector-valued mean function. The kernel function $\mathbf{k}$ is \textit{matrix}-valued, hence $\mathbf{k}:\mathbb{R}^d\times \mathbb{R}^d \rightarrow \mathbb{R}^D\times \mathbb{R}^D$, where the entry $(\mathbf{k}(\mathbf{x},\mathbf{x}^\prime))_{j,j^\prime}$ gives the covariance between the $j$th and $j^\prime$th dimension of the output. While many different choices are possible, see \cite{8187598}, we choose in this work the kernel that has been introduced in \cite{GuZa_10.5555/2981562.2981582} as \textit{multi-task} kernel. More precisely, see \cite{8187598}, this is a \textit{separable} kernel of type
$$(\mathbf{k}(\mathbf{x},\mathbf{x}^\prime))_{j,j^\prime} = k(\mathbf{x},\mathbf{x}^\prime)k_T(j,j^\prime)\,,$$
where we use a standard scalar kernel $k$, i.e.~here the Gaussian kernel, to describe the covariance in the data, and the \textit{intertask covariance} $k_T$ to capture the covariance between the output dimensions. In \cite{GuZa_10.5555/2981562.2981582}, $k_T$ is simply a $D\times D$ matrix with entries that are treated as hyperparameters. To reduce the number of to be identified hyperparameters, \cite{GuZa_10.5555/2981562.2981582} approximates this matrix is further by a low-rank approximation. We specifically use a rank-1 matrix, i.e.~we have
$$k_T(j,j^\prime) = (\mathbf{v} \mathbf{v}^\top)_{j,j^\prime}$$
and need to estimate the $D$-dimensional vector $\mathbf{v}$ as hyperparameter. This hyperparameter and any hyperparameter for the scalar kernel are collected in a set $\Phi$.

Similar to the single-valued case, we obtain
$$\mathbf{z}(\mathbf{X}) \sim \mathcal{N}(\mathbf{m}(\mathbf{X}),\mathbf{k}(\mathbf{X},\mathbf{X}))\,,$$
where the $N_T\cdot D$-dimensional mean vector $\mathbf{m}(\mathbf{X})$ concatenates the vector-valued outputs as done for $\mathbf{z}(\mathbf{X})$. The $N_T\cdot D \times N_T\cdot D$ covariance matrix $\mathbf{k}(\mathbf{X},\mathbf{X})$ is given by
\begin{equation*}
\mathbf{k}(\mathbf{X},\mathbf{X}) = \left[\begin{array}{ccc}
(\mathbf{k}(\mathbf{X},\mathbf{X}))_{1,1} & \cdots & (\mathbf{k}(\mathbf{X},\mathbf{X}))_{1,D}\\
\vdots & \ddots & \vdots\\
(\mathbf{k}(\mathbf{X},\mathbf{X}))_{D,1} & \cdots & (\mathbf{k}(\mathbf{X},\mathbf{X}))_{D,D}\\
\end{array}\right]
\end{equation*}
As before, a zero mean is assumed, giving $\mathbf{z}(\mathbf{X}) \sim \mathcal{N}(\mathbf{0},\mathbf{k}(\mathbf{X},\mathbf{X}))$.

When making predictions, we consider again $N_{eval}$ evaluation points $\mathbf{X}_\ast = \{\mathbf{x}_{\ast,1}, \ldots \mathbf{x}_{\ast,N_{eval}}\}$, giving the $N_{eval}\cdot D$-dimensional vector $\mathbf{z}(\mathbf{X}_\ast)$. We immediately consider the noisy case with zero mean, where the Gaussian noise model 
$$\mathbf{y}_i = \mathbf{z}(\mathbf{x}_i) + {\boldsymbol\epsilon}_i, \quad \mbox{with}\ {\boldsymbol\epsilon}_i \sim\mathcal{N}(\mathbf{0},\Sigma),  \quad \mbox{for}\ i=1,\ldots,N_T\,,$$
and noise covariance $\Sigma = \mbox{diag}(\sigma_1^2,\ldots, \sigma_D^2)$, is applied.
The observed data $\mathbf{y}_i$ conditioned to the GP $\mathbf{z}$, the input $\mathbf{x}_i$ and the noise covariance $\Sigma$ hence is distributed as
$$\mathbf{y}_i | \mathbf{z},\mathbf{x}_i,\Sigma,\Phi \sim \mathcal{N}(\mathbf{z}(\mathbf{x}_i),\Sigma)\,,$$
where $\Phi$ collects all potential hyperparameters of the kernel, as indicated above. For the vector of observed outputs $\mathbf{Y}$, the likelihood is given by
\begin{equation}\mathbf{Y} | \mathbf{z},\mathbf{X},\Sigma,\Phi \sim \mathcal{N}(\mathbf{z}(\mathbf{X}),\mathbf{\Sigma})\,,
\end{equation}
with $\mathbf{\Sigma} = \Sigma \otimes I_{N_T}$.

It can be shown that then the joint distribution of the training outputs ${Y}$ and the predicted outputs ${Y}_\ast$ is (with zero mean)
$$\left[\begin{array}{c} \mathbf{Y}\\\mathbf{Y}_\ast
\end{array}\right]
 \sim \mathcal{N}\left(
\mathbf{0},
\left[\begin{array}{cc} \mathbf{k}(\mathbf{X},\mathbf{X}) + \mathbf{\Sigma} & \mathbf{k}(\mathbf{X},\mathbf{X}_\ast)\\
\mathbf{k}(\mathbf{X}_\ast,\mathbf{X}) & \mathbf{k}(\mathbf{X}_\ast, \mathbf{X}_\ast)
\end{array}\right]
 \right)\,,
$$
with, e.g., $\mathbf{k}(\mathbf{X},\mathbf{X}_\ast)$ the $N_T\cdot D \times N_{eval}\cdot D$ kernel matrix combining evaluations on the training and the evaluation inputs. Condition finally gives 
\begin{align*}
\mathbf{Y}_\ast| \mathbf{z}, \mathbf{X}_\ast,\mathbf{X}, \Sigma, \Phi \sim \mathcal{N}( &\mathbf{k}(\mathbf{X}_\ast,\mathbf{X}) [\mathbf{k}(\mathbf{X},\mathbf{X})+ \mathbf{\Sigma}]^{-1} \mathbf{Y}, \\
& \mathbf{k}(\mathbf{X}_\ast,\mathbf{X}_\ast) - \mathbf{k}(\mathbf{X}_\ast,\mathbf{X})  [\mathbf{k}(\mathbf{X},\mathbf{X})+ \mathbf{\Sigma}]^{-1} \mathbf{k}(\mathbf{X},\mathbf{X}_\ast))\,.
\end{align*}
The hyperparameters can be found by maximum likelihood estimation with the log marginal likelihood, which is given, following \cite{8187598}, by
$$\log p(\mathbf{Y}|\mathbf{X},\sigma^2,\Phi) = -\frac{1}{2} \mathbf{Y}^\top [\mathbf{k}(\mathbf{X},\mathbf{X})+ \mathbf{\Sigma}]^{-1} \mathbf{Y} - \frac{1}{2} \log | \mathbf{k}(\mathbf{X},\mathbf{X})+ \mathbf{\Sigma} | - \frac{N_TD}{2} \log 2\pi\,.$$
Estimation of vector-valued GPs, when including the hyperparameter optimization, is very computaionally challenging, with $O(N_{opt} N_T^3 D^3)$ operations.

\subsection{pH-DAE identification via MT-GP regression}
As anticipated in Section~\ref{sec:identificationOfPhsDaeSystems}, the main idea behind the GP-based identification of pH-DAEs lies in the search for an effort function $\mathbf{z}(\mathbf{\mathbf{x}})$ that fits the modified problem
\begin{align*}
    E(\mathbf{x}) \cdot \dot{\mathbf{x}} - B(\mathbf{x}) \mathbf{u} = (J(\mathbf{x})-  R(\mathbf{x}))  \mathbf{z}(\mathbf{x})
\end{align*}
for given input-to-output data pairs
$$\left\{ (\mathbf{x}(t_i), E(\mathbf{x}(t_i)) \dot{\mathbf{x}}(t_i) - B(\mathbf{x}(t_i)) \mathbf{u}(t_i)) \right\}_{i=1}^{N_T}\,.$$
As before, one defines the inputs $\mathbf{x}_i := \mathbf{x}(t_i)$, $i=1,\ldots,N_T$ and collects them in $\mathbf{X}$. More importantly the output training data is defined as
$$\mathbf{y}_i := E(\mathbf{x}(t_i)) \dot{\mathbf{x}}(t_i) - B(\mathbf{x}(t_i)) \mathbf{u}(t_i))$$
and collected in a vector of outputs $\mathbf{Y}$.
While, with this data, one cannot learn the effort function directly, one can still learn its linear transformation $(J(\mathbf{x})-  R(\mathbf{x}))  \mathbf{z}(\mathbf{x})$. As discussed in \cite{Rasmussen2005}, and applied to the setting of pH-ODEs in \cite{GuZa_9992733}, a linearly transformed GP is again a GP. This gives rise to a modified regression problem, where the \textit{linearly transformed} function $\mathbf{z}_{JR}(\mathbf{x}):= (J(\mathbf{x})-R(\mathbf{x}))\mathbf{z}(\mathbf{x})$ is defined and replaced by a vector-valued Gaussian process 
$$\mathbf{z}_{JR} \sim \mathcal{GP}(\mathbf{0},\mathbf{k}_{JR})\,.$$
In this construction, $\mathbf{k}_{JR}$ is the linearly transformed kernel
\begin{equation}\label{eq:linearlyTransformedKernel}\mathbf{k}_{JR}(\mathbf{x},\mathbf{x}^\prime)= (J(\mathbf{x})-R(\mathbf{x}))\mathbf{k}(\mathbf{x},\mathbf{x}^\prime)(J(\mathbf{x})-R(\mathbf{x}))^\top\,.\end{equation}
All ideas from above identically carry over to the linearly transformed kernel, immediately giving
\begin{align*}
\mathbf{Y}_{\ast}| \mathbf{z}_{JR}, \mathbf{X}_\ast,\mathbf{X}, \Sigma, \Phi \sim \mathcal{N}( &\mathbf{k}_{JR}(\mathbf{X}_\ast,\mathbf{X}) [\mathbf{k}_{JR}(\mathbf{X},\mathbf{X})+ \mathbf{\Sigma}]^{-1} \mathbf{Y}, \\
& \mathbf{k}_{JR}(\mathbf{X}_\ast,\mathbf{X}_\ast) - \mathbf{k}_{JR}(\mathbf{X}_\ast,\mathbf{X})  [\mathbf{k}_{JR}(\mathbf{X},\mathbf{X})+ \mathbf{\Sigma}]^{-1} \mathbf{k}_{JR}(\mathbf{X},\mathbf{X}_\ast))\,.
\end{align*}
Hyperparameter optimization is done using the log marginal likelihood
$$\log p(\mathbf{Y}|\mathbf{X},\sigma^2,\phi) = -\frac{1}{2} \mathbf{Y}^\top [\mathbf{k}_{JR}(\mathbf{X},\mathbf{X})+ \mathbf{\Sigma}]^{-1} \mathbf{Y} - \frac{1}{2} \log | \mathbf{k}_{JR}(\mathbf{X},\mathbf{X})+ \mathbf{\Sigma} | - \frac{N_TD}{2} \log 2\pi\,.$$
The only step that is missing is to recover $\mathbf{z}(\mathbf{x})$ from the learned $\mathbf{z}_{JR}(\mathbf{x})$. Similar to an idea in \cite{GuZa_9992733}, this is solved by conditioning. Hence, we derive for the joint distribution of the predicted outputs $\mathbf{Y}_\ast$ and the to be found effort function outputs $\mathbf{z}(\mathbf{X}_\ast)$
$$\left[\begin{array}{c} \mathbf{Y}_\ast\\ \mathbf{z}(\mathbf{X}_\ast)
\end{array}\right]
 \sim \mathcal{N}\left(
\mathbf{0},
\left[\begin{array}{cc} \mathbf{k}_{JR}(\mathbf{X}_\ast,\mathbf{X}_\ast) & \mathbf{k}_{JR,I_{D}}(\mathbf{X}_\ast,\mathbf{X}_\ast)\\
\mathbf{k}_{I_D,JR}(\mathbf{X}_\ast,\mathbf{X}_\ast) & \mathbf{k}(\mathbf{X}_\ast, \mathbf{X}_\ast)
\end{array}\right]
 \right)\,,
$$
with the $D$-dimensional identity matrix $I_D$ in ``mixing'' kernels 
$$\mathbf{k}_{JR,I_D}(\mathbf{x},\mathbf{x}^\prime)= (J(\mathbf{x})-R(\mathbf{x}))\mathbf{k}(\mathbf{x},\mathbf{x}^\prime)(I_D)^\top = (J(\mathbf{x})-R(\mathbf{x}))\mathbf{k}(\mathbf{x},\mathbf{x}^\prime)\,,$$
$$\mathbf{k}_{I_D,JR}(\mathbf{x},\mathbf{x}^\prime)= I_D\mathbf{k}(\mathbf{x},\mathbf{x}^\prime)(J(\mathbf{x})-R(\mathbf{x})) = \mathbf{k}(\mathbf{x},\mathbf{x}^\prime)(J(\mathbf{x})-R(\mathbf{x}))^\top\,.$$
Then, the necessary mean of the Gaussian process for the effort function is found via conditioning to be
$$\overline{\mathbf{z}(\mathbf{X}_\ast)} = \mathbf{k}_{I_D,JR}(\mathbf{X}_\ast,\mathbf{X}_\ast) \mathbf{k}_{JR}(\mathbf{X}_\ast,\mathbf{X}_\ast)^{-1} \mathbf{Y}_\ast\,.$$
To summarize, the algorithm to identify the effort function for given data $\mathcal{T}=\{(\mathbf{x}(t_i),\mathbf{u}(t_i))\}_{i=1}^{N_T}$ is as follows:
\begin{enumerate}
    \item Construct the derivative data $\{\dot{\mathbf{x}}(t_i)\}_{i=1}^{N_T}$
    \item Assemble the set of inputs $\mathbf{X}$ with $\mathbf{x}_i := \mathbf{x}(t_i)$, $i=1,\ldots,N_T$ and the vector of outputs $\mathbf{Y}$ with $\mathbf{y}_i := E(\mathbf{x}(t_i)) \dot{\mathbf{x}}(t_i) - B(\mathbf{x}(t_i)) \mathbf{u}(t_i))$.
    \item Select evaluation points $\mathbf{X}_\ast$.
    \item Optimize hyperparameters using the log marginal likelihood $\log p(\mathbf{Y}|\mathbf{X},\sigma^2,\phi)$.
    \item Derive (via conditioning) the mean of the posterior $\overline{\mathbf{Y}_{\ast}| \mathbf{z}_{JR}, \mathbf{X}_\ast,\mathbf{X}, \Sigma, \Phi} = \mathbf{k}_{JR}(\mathbf{X}_\ast,\mathbf{X}) [\mathbf{k}_{JR}(\mathbf{X},\mathbf{X})+ \mathbf{\Sigma}]^{-1} \mathbf{Y}$, i.e. the predictor for the output data.
    \item Estimate the mean effort function $\overline{\mathbf{z}(\mathbf{X}_\ast)} = \mathbf{k}_{I_D,JR}(\mathbf{X}_\ast,\mathbf{X}_\ast) \mathbf{k}_{JR}(\mathbf{X}_\ast,\mathbf{X}_\ast)^{-1} \mathbf{Y}_\ast$.
\end{enumerate}
\subsection{Implementation details}
The implementation of the identification of pH-DAEs largely relies on the use of the GPyTorch package \cite{10.5555/3327757.3327857} in version 1.11 as well as the packages PyTorch 2.0.1 and NumPy 1.26.0. Visualizations are done using Matplotlib 3.8.0. In contrast to the default of the GPyTorch package, all calculations are carried out in double precision. To construct the derivative data, we build for each dimension of the data $\{\mathbf{x}(t_i)\}_{i=1}^{N_T}$ a Gaussian process without approximations (\texttt{ExactGPModel}). The derivative data is then obtained by finding the exact derivatives of the Gaussian process using standard techniques of PyTorch.

One challenge is the construction of the linearly transformed multi-task Gaussian process $\mathbf{z}_{JR} \sim \mathcal{GP}(\mathbf{0},\mathbf{k}_{JR})$. Here, we start from the class \texttt{MultitaskKernel}, which is already present in GPyTorch, and derive a new class. In that class, the application of the kernel, i.e.~the \texttt{forward} method, is implemented by evaluating the matrix-valued multitask kernel $\mathbf{k}(\mathbf{x},\mathbf{x}^\prime)$ from the parent class and then updating it according to equation~\eqref{eq:linearlyTransformedKernel}. For convenience, we also add methods to evaluate the kernels $\mathbf{k}_{JR,I_D}$ and $\mathbf{k}_{I_D,JR}$. Similarly, we derive a new mean class for the linearly transformed Gaussian process. The constructed GP model is then derived from an \texttt{ExactGP}, hence we do not apply any approximation techniques present in GPyTorch. Moreover, we can reuse the existing \texttt{MultitaskGaussianLikelihood}. A manual construction of the log marginal likelihood is not necessary, as GPyTorch does this automatically. Similarly, the minimization of the likelihood is done using standard techniques and the ADAM optimizer present in GPyTorch. The same holds for the estimation of the mean posterior $\overline{\mathbf{Y}_{\ast}| \mathbf{z}_{JR}, \mathbf{X}_\ast,\mathbf{X}, \Sigma, \Phi}$.

To calculate the mean effort function $\overline{\mathbf{z}(\mathbf{X}_\ast)}$, we extract the kernel coefficient vector ${\boldsymbol\alpha}= \mathbf{k}_{JR}(\mathbf{X}_\ast,\mathbf{X}_\ast)^{-1} \mathbf{Y}_\ast$ via \texttt{model.prediction\_strategy.mean\_cache} from the trained linearly transformed multi-task GP model and apply to it the properly evaluated, previously implemented, kernel $\mathbf{k}_{I_D,JR}$.


\section{Numerical results}
\label{sec:numericalResults}
\subsection{Model systems}

We consider two model systems from electrical network design   and constrained multibody systems modeling.

\subsubsection{Electrical network test example}
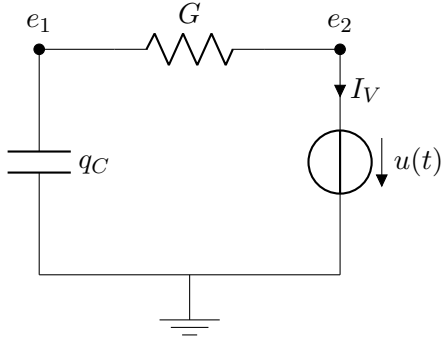
\begin{figure}
 \begin{center}
 \input{circuit.tex}
    \end{center}\vspace*{-3em}
    \caption{Network consisting of a nonlinear capacitance at node $e_1$ with charge function $q_C=u_C^2/2$, a linear resistor with conductance $G$ between nodes $e_1$ and $e_2$ and a voltage source $u(t)$ at node $e_2$.}
    \label{fig:circuit}
\end{figure}

The first model is given by an electrical network  model. It consists of a circuit with two nodes and three 
network elements, see Figure~\ref{fig:circuit}. It has a 
    a capacitance at node $e_1$ with charge function $q_C=q(u_C)=u_C^2/2$,
    a linear resistor with resistance $R$ between nodes $e_1$ and $e_2$ and
    a time-dependent voltage source with input $u(t)$ at node $e_2$ and current $I_V$ through the voltage source.

Using the charge-conserving modelling approach defined in~\cite{gubajare21} yields
a semi-explicit index-1 pH-DAE with 
$\mathbf{x}=(x_1, x_2, x_3)^T=(q_C, u_2, I_V)^\top$ and nonlinear function 
$\mathbf{z}(\mathbf{x})=(\sqrt{2 q_C},u_2, I_V)^\top$:
\begin{equation}
E \cdot \mathbf{x} =   (J-R) \mathbf{z}(\mathbf{x}) + B\mathbf{u}, \quad \mathbf{y} = B^\top \mathbf{z}(\mathbf{x})
\end{equation}
with system matrices
$$
E=\begin{bmatrix}
    1 & 0 & 0 \\ 0& 0 & 0  \\ 0 & 0 &0 
\end{bmatrix}, \quad
J=\begin{bmatrix}
    0 & 0 & 0  \\ 0 & 0 & 1 \\ 0  & -1& 0
\end{bmatrix}, \quad
R=\begin{bmatrix}
    G
    &  -G 
    & 0  \\  -G 
    & G 
    & 0 \\  0 &0 & 0
\end{bmatrix}, \quad
B= \begin{bmatrix}
    0 \\ 0  \\ 1
\end{bmatrix}
$$
and Hamiltonian
$$
H(\mathbf{x})= \frac{2}{3} \sqrt{2} x_1^{\frac{3}{2}},$$
which only depends on $x_1$. Note that $J-R$ is regular for $G\neq 0$.
One easily verifies the compatibility condition $ E^\top \mathbf{z}(\mathbf{x}) = \nabla H(\mathbf{x})$, which
reads in this case:
$$
\begin{bmatrix}
    z_1(\mathbf{x}) 
\end{bmatrix}
= \nabla_{x_1} H(x_1), \quad
\nabla_{x_2,x_3} H(x_1) =0.
$$

\subsubsection{Constrained multibody system test example}
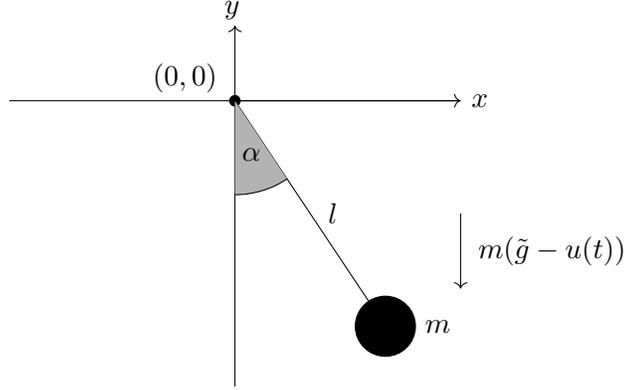
\begin{figure}[h]
    \centering \hspace*{2.5cm}
   \input{pendulum.tex}
    \caption{Mathematical pendulum in the $(x,y)$ plane with concentrated mass $m$, length $l$ and gravitational constant $\tilde g$ and applied force defined by the input $u(t)$. The position of the mass is given by $\mathbf{q}=(x,y)^\top$, the deflection angle $\alpha$ is defined by $\arctan(-x/y).$}
    \label{fig:pendulum}
\end{figure}

The equation of motion for a constrained multibody system, described by the augmented Hamiltonian 
$$
H(\mathbf{q},\mathbf{p},\boldsymbol{\lambda}):=
\frac{1}{2} \mathbf{p}^\top M^{-1} \mathbf{p} + U(\mathbf{q}) - g(\mathbf{q})^\top \boldsymbol{\lambda}
$$
with constraint $0=g(\mathbf{q})$, is given by the Port-Hamiltonian DAE systen for  $\mathbf{x}=(\mathbf{q},\mathbf{p},\boldsymbol{\lambda})$ as
\begin{align}
\label{ex.mech}
\mbox{diag}(I,I,0) 
    \mathbf{\dot x}
  & = 
    \left( J-R
    \right)
    \mathbf{z}(\mathbf{x}) + B \mathbf{u}
\end{align}
with
\begin{align*}
       J&= \begin{bmatrix}
       0 & I & 0 \\ -I & 0 & 0 \\ 0 & 0 & 0
    \end{bmatrix}, \,\,
    R = \begin{bmatrix}
        0 & 0 & 0 \\ 0 & 0 & 0 \\ 0 & 0 &I
    \end{bmatrix}, \,\,
    B = \begin{bmatrix}
        0 \\ \mathbf{b} \\ 0 
    \end{bmatrix}, \,\,
    \mathbf{z}(\mathbf{x}) = \begin{bmatrix}
            H_\mathbf{q}(\mathbf{q},\mathbf{p},\boldsymbol{\lambda})
 \\
H_\mathbf{p}(\mathbf{q},\mathbf{p},\boldsymbol{\lambda}) 
 \\
 H_\lambda(\mathbf{q},\mathbf{p},\boldsymbol{\lambda}) 
    \end{bmatrix}=
    \begin{bmatrix}
        \frac{\partial U(\mathbf{q})}{\partial \mathbf{q}} - \frac{\partial g(\mathbf{q})}{\partial \mathbf{q}}^\top \boldsymbol{\lambda} \\
        M^{-1} \mathbf{p} \\ -g(\mathbf{q})
    \end{bmatrix},
\end{align*}
if an external force $B \mathbf{u}(t)$ is applied to the system.
The compatibility condition diag$(I,I,0)^\top \mathbf{z}(\mathbf{x}) = \nabla H(\mathbf{x})$ holds due to $g(\mathbf{q})=0$.
\if
For the simple pendulum in the plane of length $l$, mass $m$ and the gravitational constant $\tilde g$~\cite{Simeon_2013}, we have  
$$
M=\mbox{diag} (m,m,m l/12), \, 
U(q)=\begin{bmatrix}
    0 \\ g m q_2 \\ 0
\end{bmatrix}, \,
g(\mathbf{q})=\begin{bmatrix}
    q_1-\frac{l}{2} \cos(q_3) \\ q_2 - \frac{l}{2} \sin(q_3)
\end{bmatrix}, \,
b= \begin{bmatrix}
    0 \\ 1 \\ 0
\end{bmatrix},
$$
if an external force  is applied to the the pendulum acting in $y$-direction.
\fi
For the simple mathematical pendulum in the plane of length $l$, mass $m$ and gravitational constant $\tilde g$~\cite{Simeon_2013} shown in Fig.~\ref{fig:pendulum} we have  
$$
M=\mbox{diag} (m,m), \, 
U(q)=\tilde{g} m q_2, \,
g(\mathbf{q})=\mathbf{q}^\top \mathbf{q} - l^2, \,
\mathbf{b}= \begin{bmatrix}
    0 \\ m 
\end{bmatrix},
$$
if an external force  is applied to the  pendulum acting in $y$-direction.
\subsection{Benchmark data sets}
The data utilized in our numerical results is generated by numerical integration from the respective test examples. For the first test example, we set $G=1$ and $u(t)=a \cdot(1+\sin(\omega t))$ and choose the 
parameters $a= 1$ and $\omega = 1$.
For these parameter choices, we solve first the scalar ODE
$$
\dot x_1 = G \cdot (u(t)-\sqrt{2 x_1}), \quad x_1(0)=2
$$
on $t \in [0, 10 \pi/ \omega]$ using the explit Euler scheme with constant step size $10 \pi/ (3000 \omega)$. 
For the derived data $\left\{
(t_i,x_1(t_i)\right\}_{i=0}^{N_T}$ with
$t_0=0$ and $t_{N_T}=10 \pi$, we then define the algebraic components $\{x_2(t_i)\}_{i=0}^{N_T}$ and $\{x_3(t_i)\}_{i=0}^{N_T}$ by
\begin{align*}
    x_2(t_i) & := u(t_i), \\
    x_3(t_i) & :=G \cdot (x_1(t_i)-x_2(t_i)).
\end{align*}
If needed, the derivative data $\{\dot x_1(t_i)\}_{i=0}^{N_T}$  can be obtained by
$$
\dot x_1(t_i) = G \cdot(u(t_i)-\sqrt{2 x_1(t_i)}).
$$

For the second test example, we set $m=l=\tilde g=1$ 
and choose the input $u(t)=\beta \tilde{g} \cos(\frac{2 \cdot \pi}{\tau}  t)$
for $\tau = 1$ and $\beta = 1$.
For these parameter choices, we solve the equation of motion in minimal coordinates given by the scalar second order ODE
$$
\ddot \alpha = -\frac{1}{l} \left( \tilde g-u(t) \right) \sin{(\alpha(t))},
\quad\alpha(0)=\frac{\pi}{4}, \, \dot \alpha(0)=0
$$
on $t \in[0,30]$ using the explicit Euler scheme with step size $h=0.01$ applied to the transformed first order system. 

For the derived data $\left\{
(t_i,\alpha(t_i)\right\}_{i=0}^{N_T}$ and  $\left\{
(t_i,\dot \alpha(t_i)\right\}_{i=0}^{N_T}$ with
$t_0=0$ and $t_{N_T}=30$, we then define the solution of the pH-DAE~\eqref{ex.mech} $\{q(t_i)\}_{i=0}^{N_T}$,  $\{p(t_i)\}_{i=0}^{N_T}$and $\{\lambda(t_i)\}_{i=0}^{N_T}$ by
\begin{align*}
    q(t_i) & := l \begin{bmatrix}
        \sin(\alpha(t_i)) \\ - \cos(\alpha(t_i))
    \end{bmatrix}, \\
    p(t_i) & := ml \dot \alpha \begin{bmatrix}
        \cos(\alpha(t_i)) \\  \sin(\alpha(t_i))
    \end{bmatrix}, \\
    \lambda(t_i) & :=\frac{m}{2l} \left( \cos(\alpha(t_i)) (u(t_i)-\tilde g) - l \dot \alpha(t_i)^2\right).
\end{align*}
If needed, the derivative data $\{\dot q(t_i)\}_{i=0}^{N_T}$ and  $\{\dot p(t_i)\}_{i=0}^{N_T}$ can be obtained by
$$
\dot q(t_i) = \frac{1}{m} p(t_i), \quad
\dot p(t_i) = \begin{pmatrix}
    0 \\ -m \tilde g
\end{pmatrix} + 2 q(t_i) \lambda(t_i) +
\begin{pmatrix}
    0 \\ m
\end{pmatrix}
u(t_i).
$$
\subsection{Benchmark setup}
In the construction of the Gaussian processes for the derivative data, we use the Gaussian kernel, where the hyperparameters are automatically determined by minimizing the respective negative log marginal likelihoods using the ADAMS optimizer with a learning rate of $0.1$. The optimizer is stopped after 200 iteration with a stagnating likelihood.
Note that the approximation of the derivatives has an impact on the prediction. It is necessary to differentiate the case of first calculating the derivatives on the full dataset and then selecting subsets of this dataset for training and the case of first extracting these subsets and only after that applying the derivative calculation on the subsets. While in the first case, a consistent (good) approximation of the derivative data is obtained, the derivatives in the second case are constructed from potentially much worse resolved datasets, leading to potentially strongly deteriorated derivative approximations. Still, the second case is the more realistic setting from a practical perspective of real data. While our main numerical error analysis of the identification method in the next subsection will work with idealized data, i.e.~the first case, we provide a separate analysis on the impact of the derivative data approximation in Section~\ref{sec:influenceOfDerivativeApproximation}. 

The main regression task using the linearly transformed multi-task GP, is also carried out with the Gaussian kernel. As before, the set of hyperparameters is automatically found with 200 steps of the ADAMS optimizer with learning rate 0.1. Errors are always reported as root mean square errors (RMSE). To test the prediction error of the constructed Gaussian processes, $N_{test}$ samples of the original dataset are randomly sub-selected and set aside as test set $\mathcal{T}_{test}$, which is kept identical throughout the full analysis. Prediction errors are then reported using \textit{learning curves}, i.e.~we report the RMSE of the model on the test set for a growing number of training samples in a double-logarithmic plot. The training samples are randomly sub-selected from the dataset $\mathcal{T} \setminus \mathcal{T}_{test}$, such that in a single test run $N_{T}=2,4,8,16,\ldots$~samples are extracted that form nested sets. The RMSE for each of the training sets then forms the data for a single learning curve. Since each learning curve is however subject to the randomization of the data selection, the process is repeated such that five such learning curves are constructed and averaged. In general, we expect the learning curves to decrease for growing number of training samples. However, it may happen that -- starting from a given amount of training data -- the model no longer improves. This can have various reasons including noise in the data, the selected regularization, etc., which are discussed, if necessary.

\subsection{Identification error analysis: Electrical network test example}
\label{sec:identificationErrorAnalysisCircuitExample}
\begin{figure}[tbh]
    \begin{center}
    \includegraphics[scale=0.59]{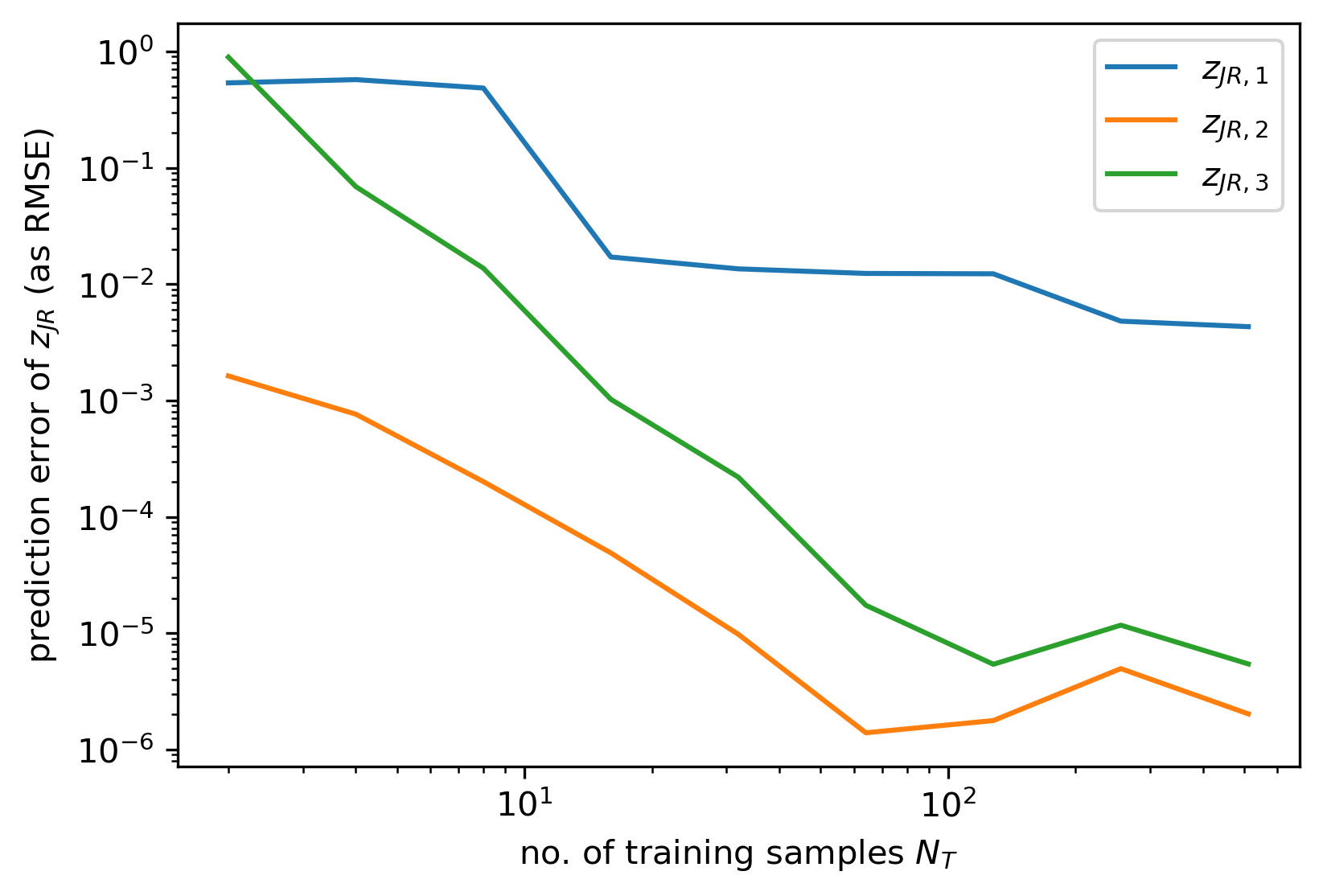}\hspace*{0.0em}
    \includegraphics[scale=0.59]{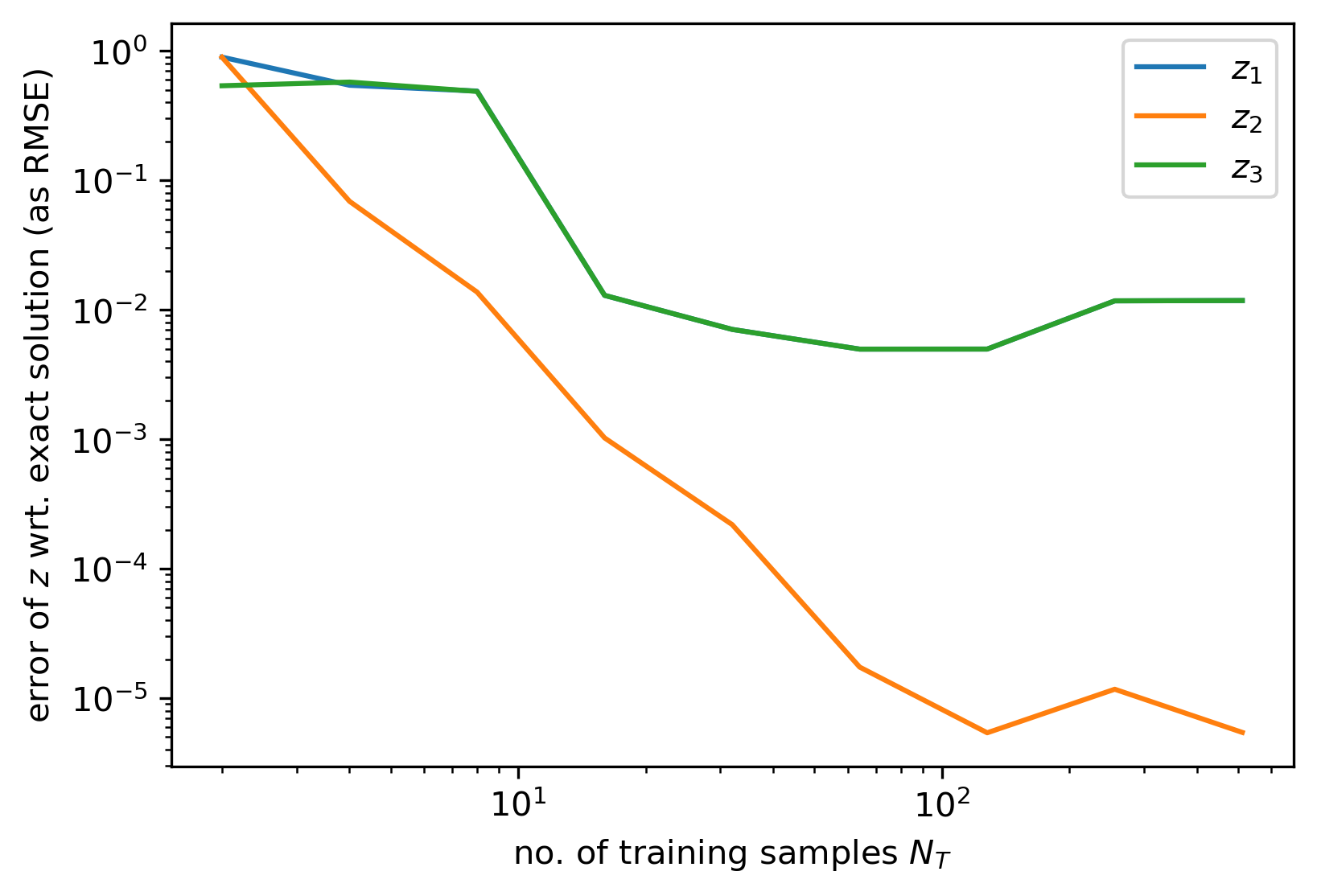}
    \end{center}
    \caption{\label{fig:learningCurveCircuit}\textit{Electrical network test example:} For growing number of training samples, the prediction error of $\mathbf{z}_{JR}$ (left) and the finally reconstructed effort function $\mathbf{z}$ (right), decay. The first component stagnates at the noise level of the data.}
\end{figure}
In the following, we study the identification error that we obtain, when using the newly introduced pH-DAE effort function identification method. The identification comprises two parts. In a first step the GP $\mathbf{z}_{JR}:=(J-R) \mathbf{z}$ is estimated, which is the main \textit{learning} task. The output samples in this electrical network test  example are $\mathbf{y}_i=(\dot{x}_1,0,-u(t))$, with $i=0,\ldots,N_T$. When analyzing this data, one observes that only the first component $z_{JR,1}(\mathbf{x})$ is dependent on data ($\dot{x}_1$) that is ``noisy'', more precisely it comes from the solution of a discretized pH-ODE, as discussed in Section~\ref{sec:identificationOfPhsDaeSystems}. The output data for the second component $z_{JR,2}(\mathbf{x})$ is constant zero, while the third 
trained component $\mathbf{z}_{JR,3}(\mathbf{x})$ only depends on the analytically given input. This has an impact on the prediction error reported for growing number of training samples as learning curve in Figure~\ref{fig:learningCurveCircuit}, on the left-hand side. While the second and third component consistently decrease until they reach an error level of about $10^{-5}$ with 128 training samples, the error in the first component already stagnates in an error range of $10^{-2}$ for only 16 samples. The stagnation of the error at the error level of $10^{-5}$ comes from the fact that optimizer selects the regularizing noise variance $\sigma^2$ in the range of $10^{-5}$ for the various runs. The surprisingly early stagnation of the first component at about $10^{-2}$ perfectly matches the numerical noise of the state space data. As a reminder, we generate the training data using the first-order accurate explicit Euler method with a step size of $10^{-2}$. This noise gets immediately translated into the prediction error of the method for the first component.

\begin{figure}[tbh]
    \begin{center}
    \includegraphics[scale=0.59]{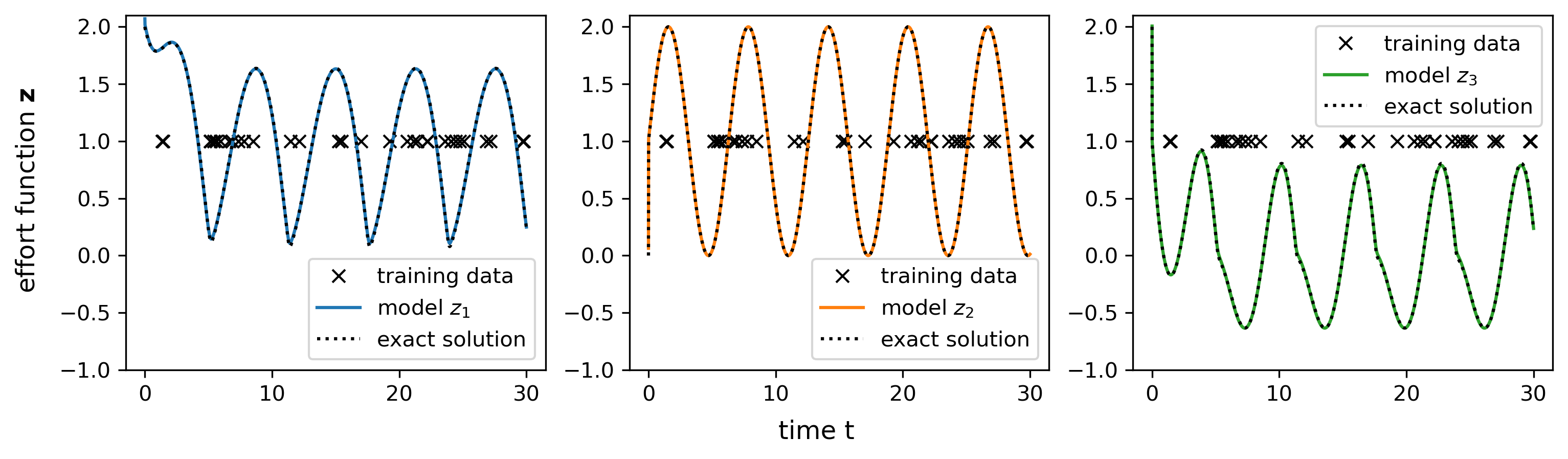}\\
    \includegraphics[scale=0.59]{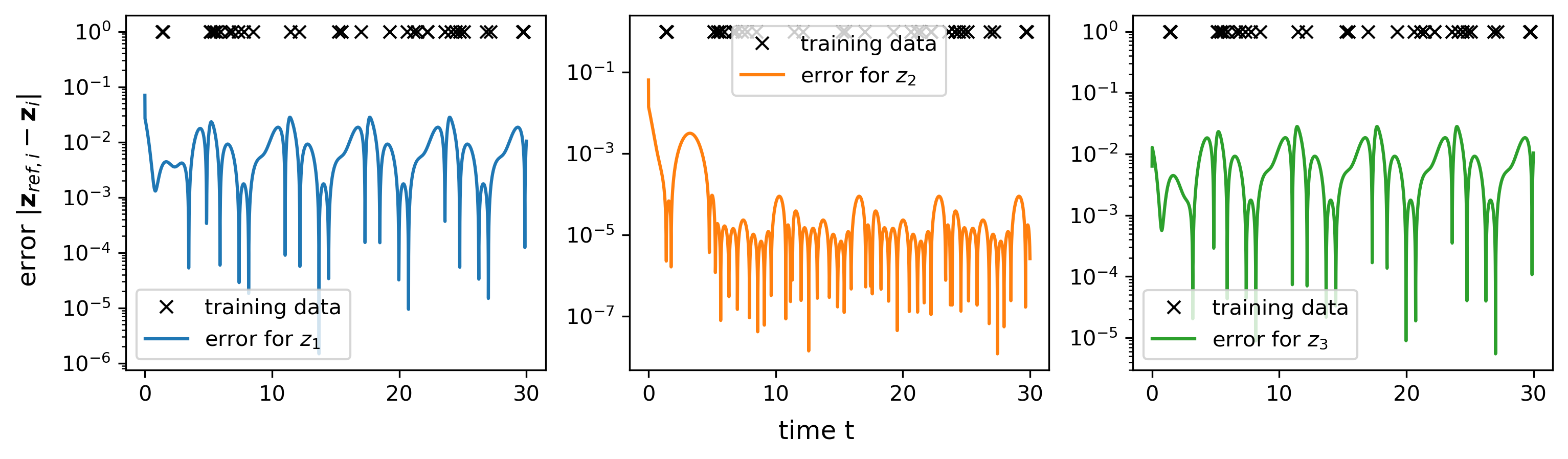}
    \end{center}\vspace*{-1.5em}
\caption{\label{fig:modelVsSolutionCircuit}\textit{Electrical network test example:} With only $N_T=32$ training samples, the evaluation of the identified $\mathbf{z}(\mathbf{x}(t))$ on the interval $t=[0,30]$ is visually indistinguishable from the exact solution (top), with an error in the range of the discretization error that occurred in the training data generation (bottom).}
\end{figure}
In a second step, the actual effort function $\mathbf{z}(\mathbf{x})$ is derived. Evaluation errors for the effort function on the test data are reported in Figure~\ref{fig:learningCurveCircuit}, on the right-hand side. The matrix $(J-R)$ couples the first two components of the effort function, while the third component of the identified function $\mathbf{z}_{JR}(\mathbf{x})$ is connected to the second component of the effort function. As a consequence, the first components of $\mathbf{z}(\mathbf{x})$ show a similar error stagnation in the range of $10^{-2}$ as the first component of $\mathbf{z}_{JR}(\mathbf{x})$, while the third component shows the exact same error behavior as 
$\mathbf{z}_{JR,2}(\mathbf{x})$, which is the expected behavior. Overall, we obtain a prediction error in the range of the discretization error, hence data error, range of the training data with only 16 training samples, which is a quite significant result.

To get an intuition for the identified effort function, we evaluate in Figure~\ref{fig:modelVsSolutionCircuit} the identified effort function trained on $N_T=32$ training samples, which are visualized by black crosses, on all training points and report both the evaluated effort function and its point-wise error with respect to the exact solution. Visually, there is now way to distinguish the model from the exact solution. The error is rather equally distributed over the full time interval. In the first and second component of the effort function, it becomes noticeable that the random selection of the training samples did not select training samples towards the beginning of the time interval. Therefore, initially, the effort function shows an increased error, which is not visible in the full RMSE due to averaging. To avoid such artifacts, one may want to enforce a more even distribution of training samples.

\begin{figure}[t]
    \begin{center}
    \includegraphics[scale=0.59]{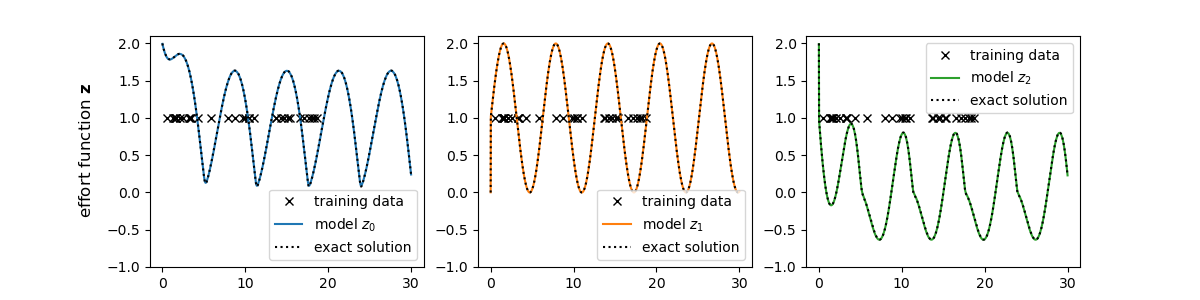}
    \end{center}
    \caption{\label{fig:modelVsSolutionExtrapolationCircuit}\textit{Electrical network test  example:} The model still properly generalizes to the time interval $t\in [20,30]$, even if the $N_T=32$ training samples are only taken from the time interval $t\in [0,20]$.}
\end{figure}

In a further study, depicted in Figure~\ref{fig:modelVsSolutionExtrapolationCircuit}, we limit the random selection of the training data only to the time interval $t\in [0,20]$ and still evaluate the model on the full available state space data. Thereby, we can observe, how well the model extrapolates using the given data. Even in this more challenging analysis, the model with $N_T=32$ training samples matches the exact solution with a similar error range as before. In fact, this is no surprising result as the effort function does not directly depend on the time, however it depends on the state of the system. By picking samples from the time interval $t\in [20,30]$, we already properly cover the state space (of this limited test case), hence the effort function gets properly resolved.

\subsection{Identification error analysis: Constrained multibody system example}
\begin{figure}[tbh]
    \begin{center}
    \includegraphics[scale=0.58]{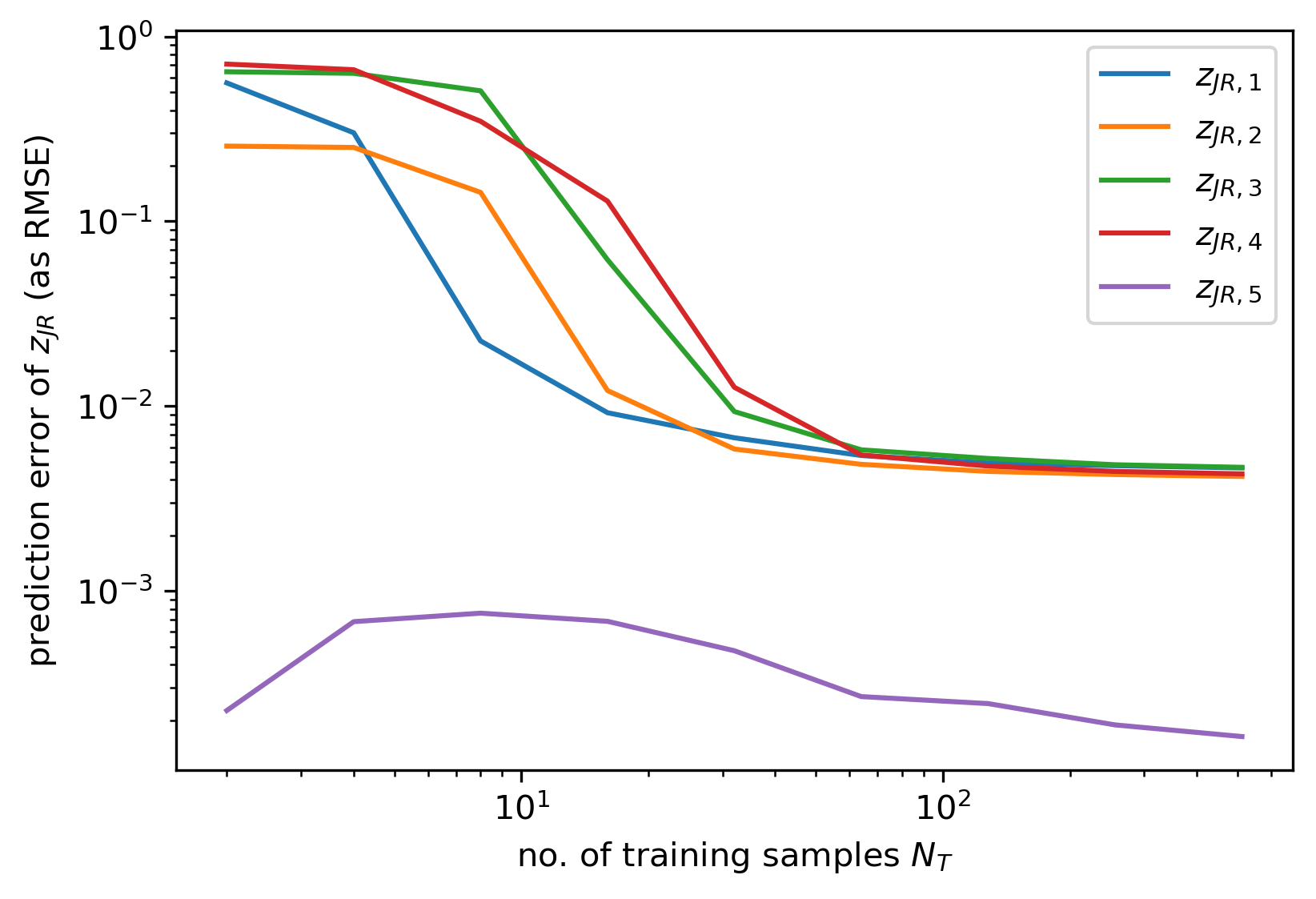}\includegraphics[scale=0.58]{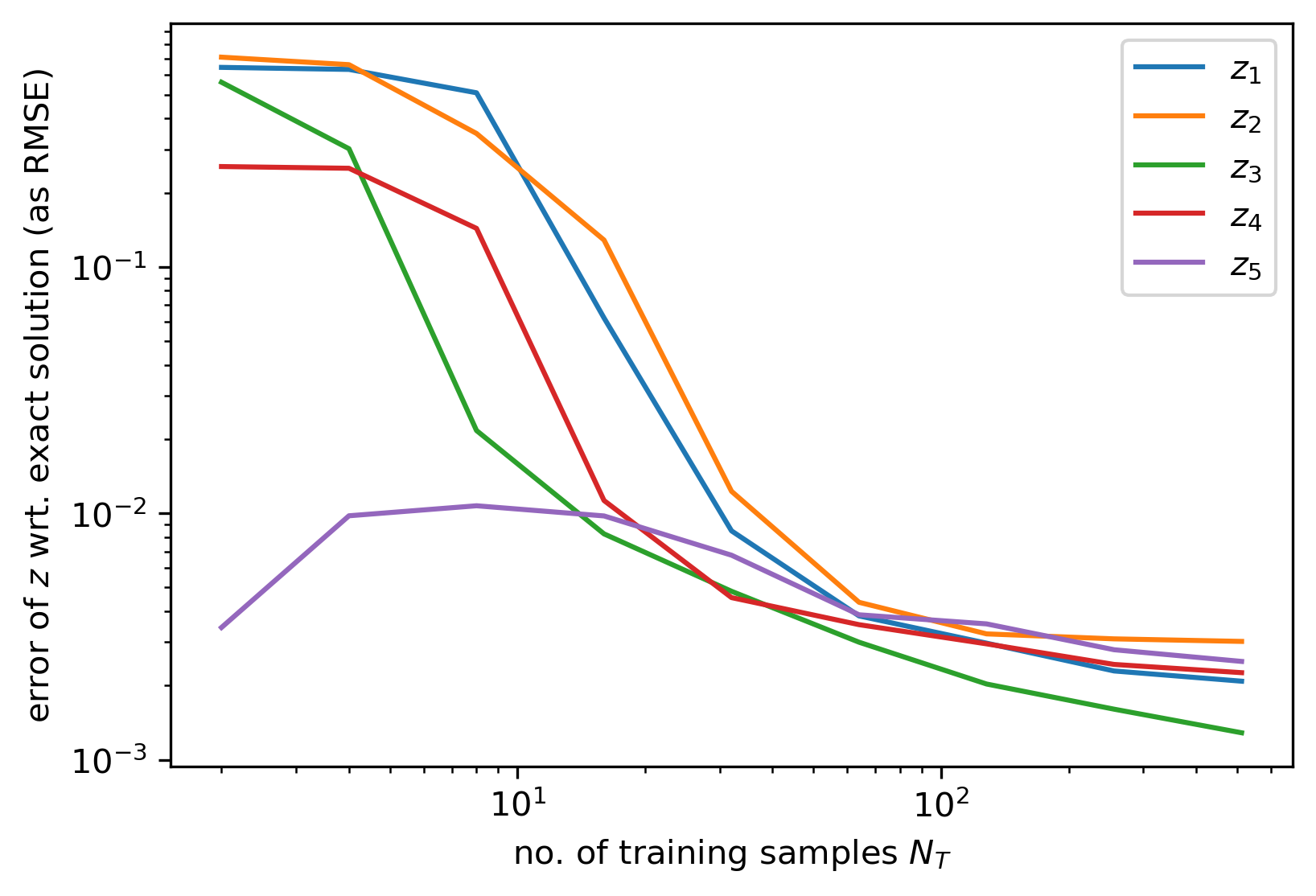}
    \end{center}
    \vspace{-1em}\caption{\label{fig:learningCurveMechanics}\textit{Constrained multibody system example:} All components of the transformed effort function $\mathbf{z}_{JR}$ \textit{(left)} and the finally identified effort function $\mathbf{z}$ \textit{(right)} reach a prediction error of at least the discretization error of the data with $N_T=64$ training samples.}
\end{figure}

We repeat the same study as for the electrical network test case for the constrained multibody system, i.e.~the pendulum, to verify that the approach is generally applicable to different applications. As a small deviation from the electrical network test case, we here approximate the derivative information for $\dot{\mathbf{x}}$ by simple finite differences. All other parameters remain identical.

In Figure~\ref{fig:learningCurveMechanics}, the learning curves for the transformed effort function $\mathbf{z}_{JR}$ and the finally identified effort function $\mathbf{z}$ are depicted. The prediction error of the actually ``trained'' transformed effort function $\mathbf{z}_{JR}$, decreases until reaching the discretization error limit for the first four components. Only the error for the fifth component reaches a lower level. The reason for this is that the first four components rely on the (discretized) state space information and its time derivative, hence are subject to the time discretization error. The fifth component is only weakly depending on this data via the input features to the model, while in fact approximating the constant zero function. This can apparently be done at higher accuracy. Overall, the discretization error level is reached with about $N_T=64$ training samples. Very similar to this is the result for the finally identified effort function, in the same Figure~\ref{fig:learningCurveMechanics}. Again, all components have reached the discretization error limit with about $N_{T}=64$ training samples.

\begin{figure}
    \begin{center}
    \includegraphics[scale=0.58]{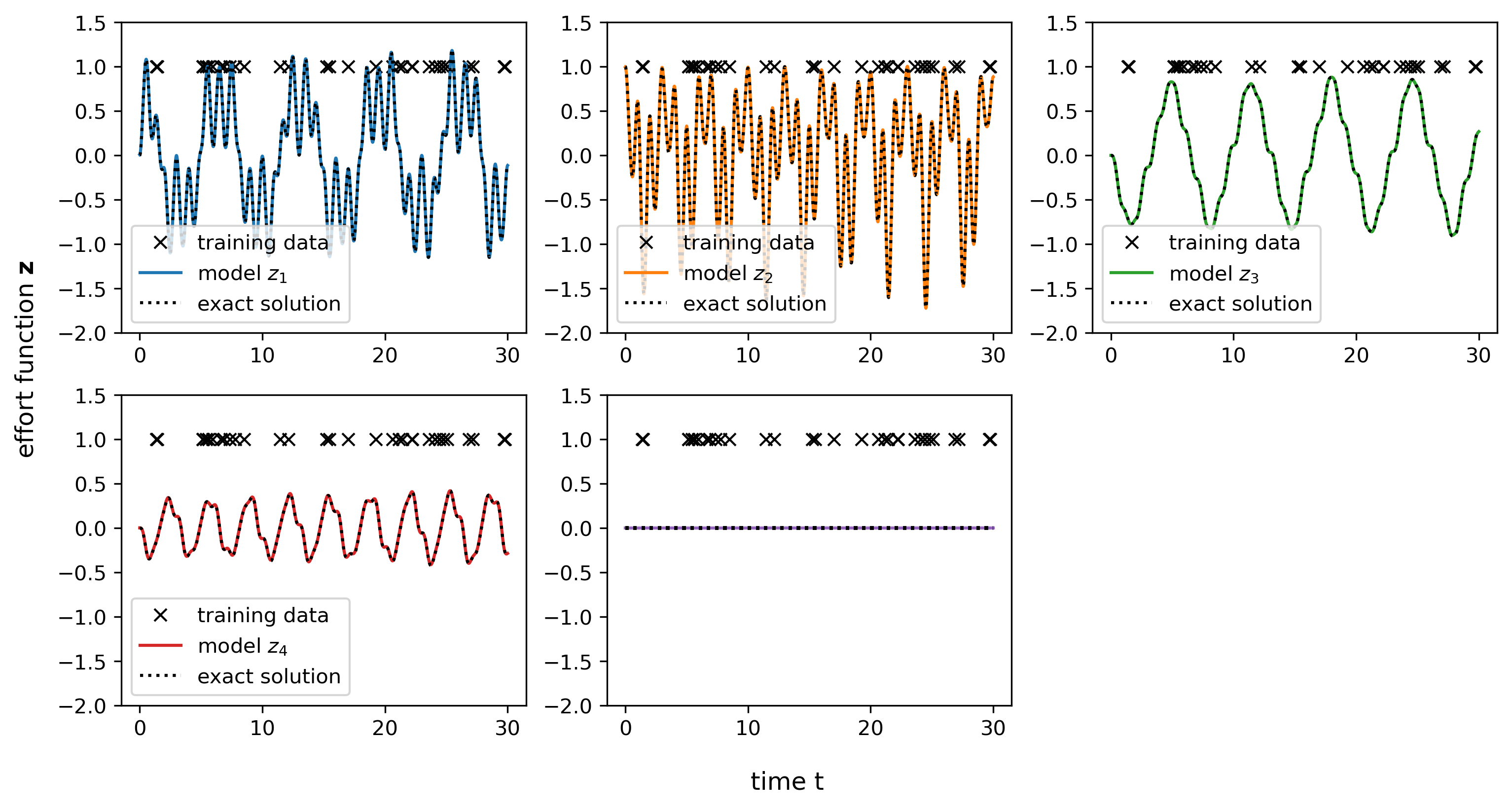}
    \includegraphics[scale=0.58]{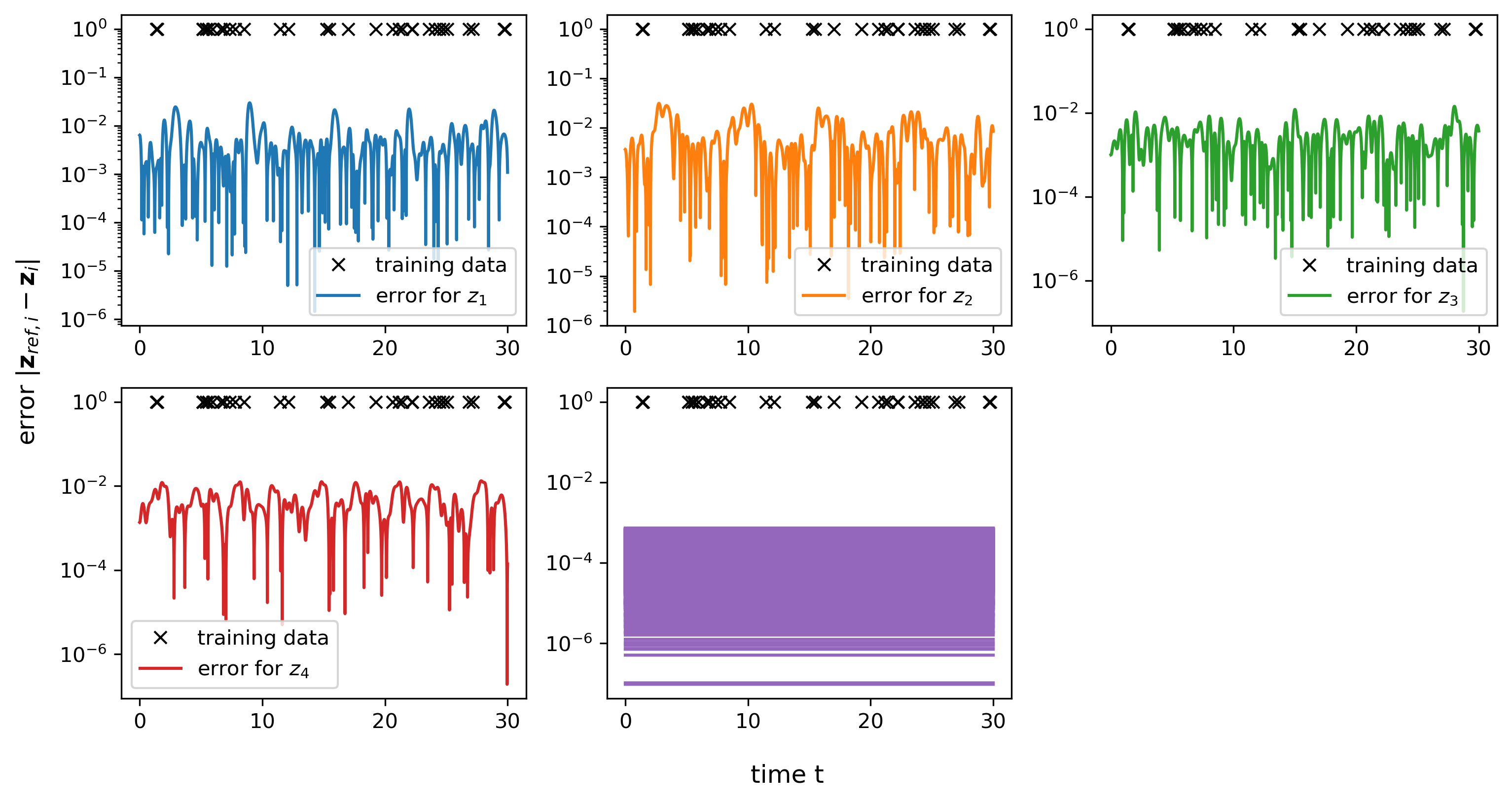}
    \end{center}
    \vspace{-1.5em}
\caption{\label{fig:modelVsSolutionMechanics}\textit{Constrained multibody system example:} Already with $N_T=32$ training samples, the model for the individual components of the effort function is visually no longer distinguishable from the exact solution \textit{(top)}. In that case, a general error level of about $10^{2}$ is reached \textit{(bottom)}.}
\end{figure}
\begin{figure}[tbh]
    \begin{center}
    \includegraphics[scale=0.58]{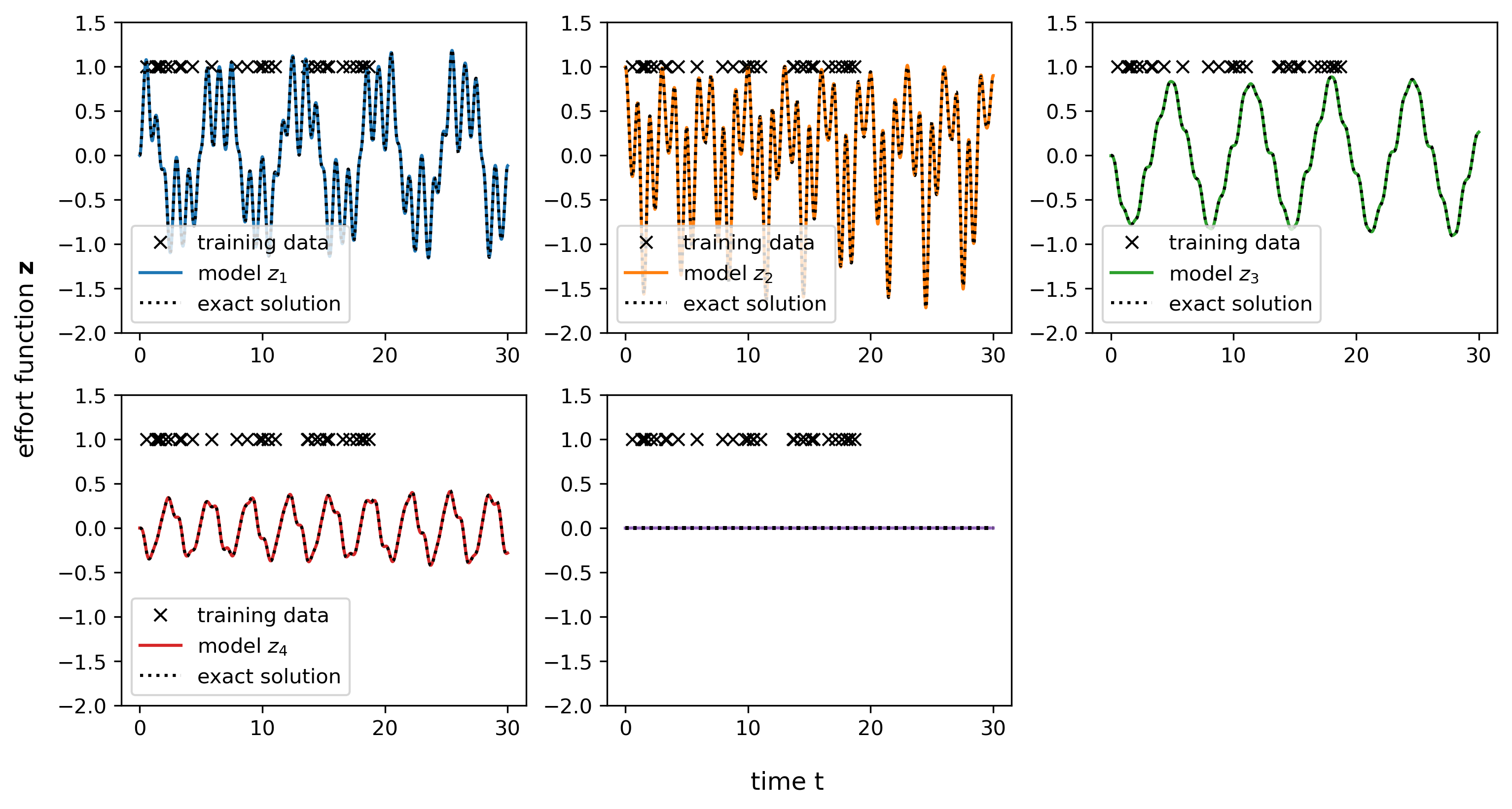}
    \end{center}
    \vspace{-1.5em}
    \caption{\label{fig:modelVsSolutionExtrapolationMechanics}\textit{Constrained multibody system example:} For $N_T=32$ training samples from the time interval $t\in [0,20]$, the identified effort function properly generlizes towards the non-observed time-interval $t\in [20,30]$}
\end{figure}

Beyond that, Figure~\ref{fig:modelVsSolutionMechanics} depicts the predicted effort function output for $N_{T}=32$. Visually, the predicted effort function output and the exact solution are full agreement, while the error for this model, depicted in the same figure, is typically in the range of the discretization error of the state space data, i.e.~$10^{-2}$. In Figure~\ref{fig:modelVsSolutionExtrapolationMechanics}, the results of the final extrapolation study are presented. With $N_T=32$ training samples taken from the time interval $t\in [0,20]$, the effort function still gets properly extrapolated towards the non-observed time-interval $t\in [20,30]$.

\subsection{Influence of derivative approximation}\label{sec:influenceOfDerivativeApproximation}
We return to the electrical network test example in order to analyze the influence of the approximation of the derivatives $\dot{\mathbf{x}}$ on the identification of the effort function. To this end, we redo the numerical experiments with learning curves from Section~\ref{sec:identificationErrorAnalysisCircuitExample}. However, this time, we compare results for (1) analytically exact derivative data, (2) derivative data generated with GPR on the full data set as used before, and (3) derivative data generated with GPR only on the individual training sets. We had earlier, in Section~\ref{sec:identificationErrorAnalysisCircuitExample}, motivated that the third case is indeed the most realistic one, however, it is expected to introduce a higher error in the derivative approximation.

\begin{figure}[tbh]
    \begin{center}
    \includegraphics[scale=0.59]{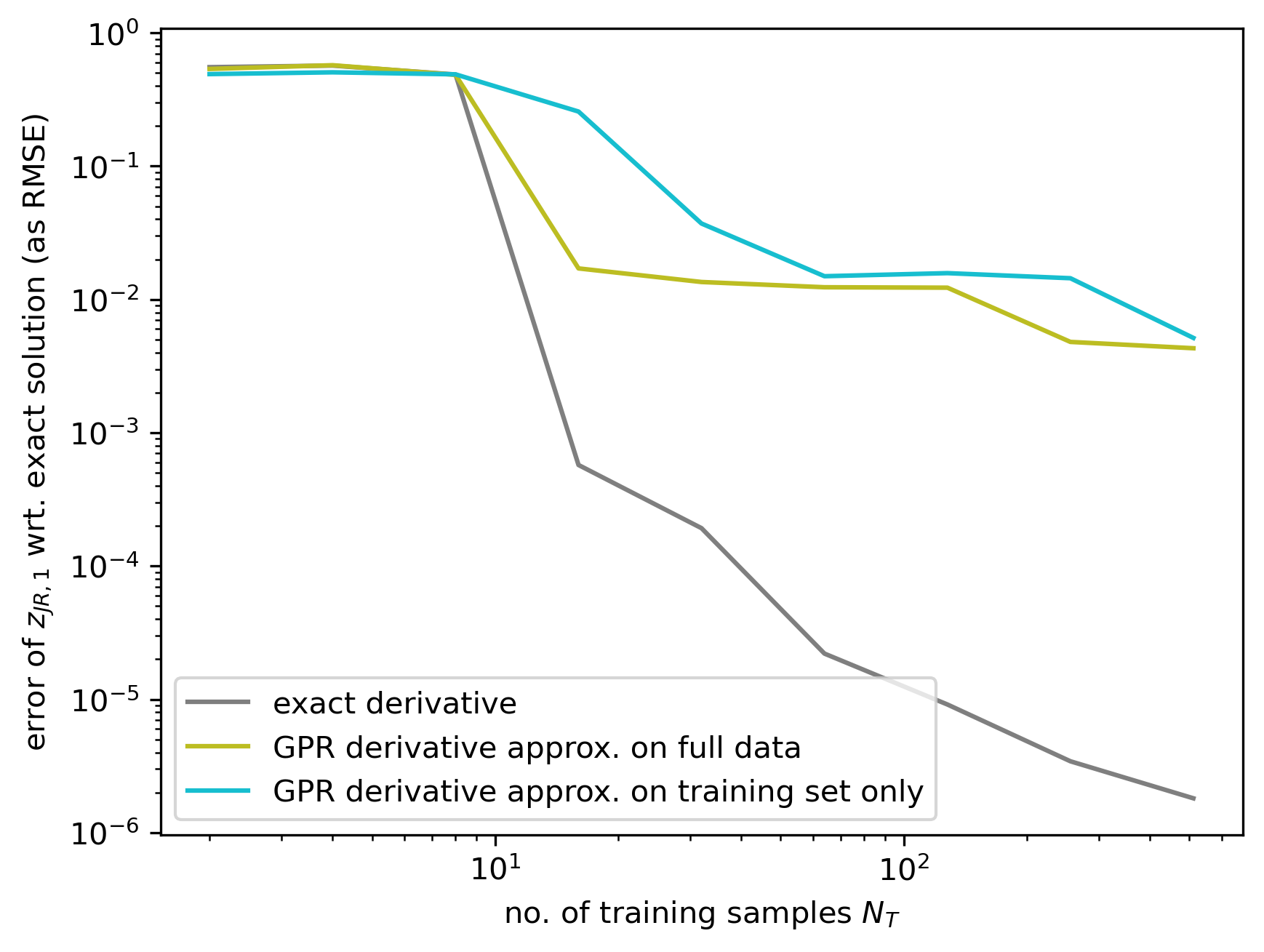}\includegraphics[scale=0.59]{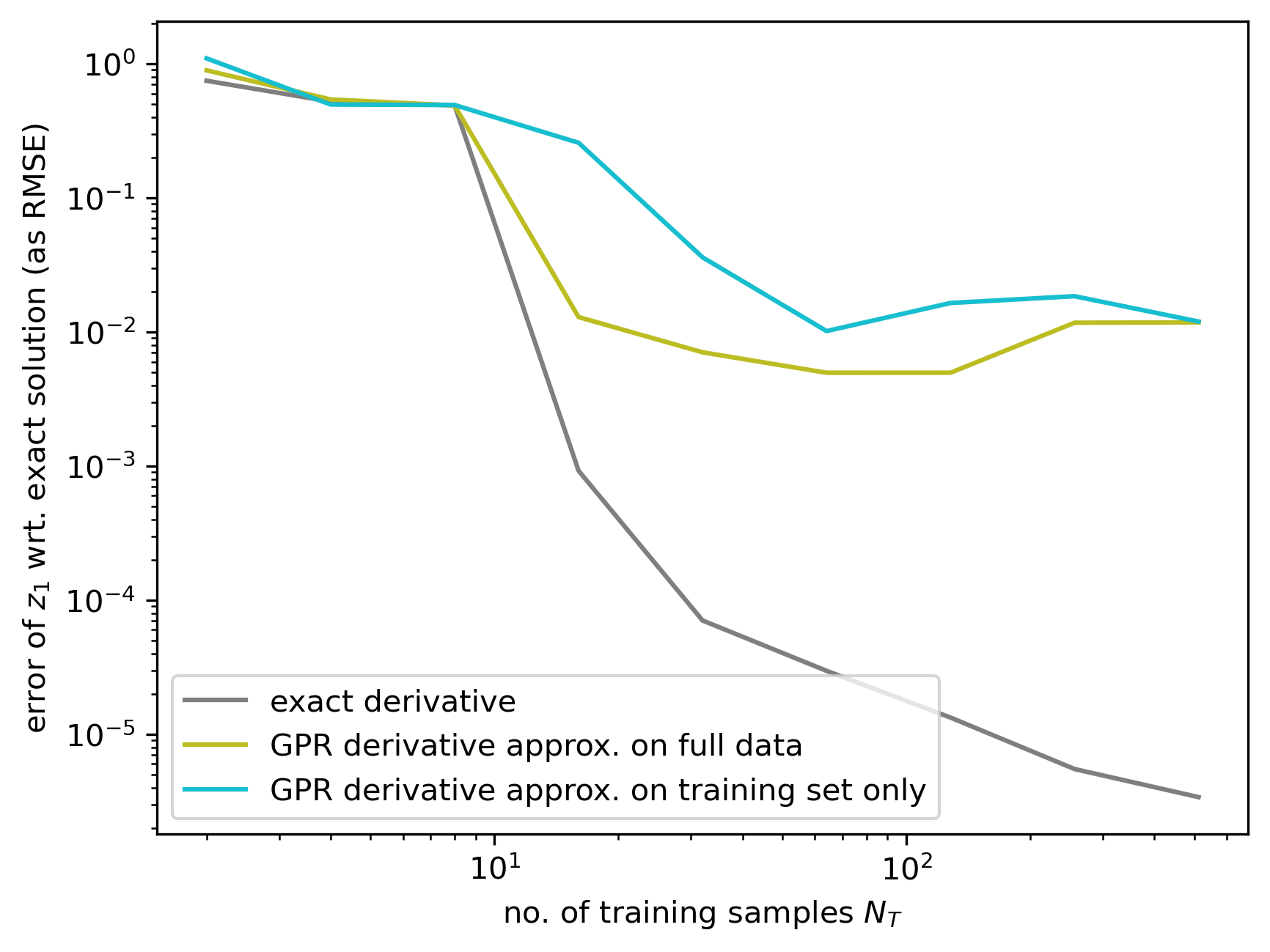}
    \end{center}
    \caption{\label{fig:derivativeApproximationExample}\textit{Electrical network test  example:} The approximation of the derivative data for $\dot{\mathbf{x}}$ influences the prediction error for $z_{JR,1}$ (left) and $z_1$ (right). Still, all approximate derivative calculations reach the discretization error of the underlying data.}
    
\end{figure}

The results for this study are depicted in Figure~\ref{fig:derivativeApproximationExample} for the first component of the transformed effort function $z_{JR,1}$ and for the first component of the finally identified effort function $z_1$.
When using exact derivative data, the error no longer stagnates at the discretization error level of $0.01$. This is an expected result, as the discretization error has no influence on the analytically derived derivative. Of higher relevance is the comparison between the use of GPR derivative approximations on the full data set and the training data set only. Indeed, even thought the derivatives are generated in the latter case from much fewer points, the prediction error, for growing number of training samples, also reaches the discretization level. This shows, that in this test case, the discretization and the prediction error are properly balanced: Both, the prediction error and the derivative approximation error for growing training set size decay in a similar way, leading to an overall converging scheme. Still, it needs to be noted that the more realistic GPR derivative approximation on only the training set requires more training for a model of similar accuracy, compared to the other GPR approximation approach.


\section{Conclusions}
In this work, we have introduced an approach for the identification of the effort function of a nonlinear pH-DAE system from input and state space data. The effort function is replaced by a  multi-task Gaussian process, which allows its identification on noisy data. In our results, we have shown that the developed approach succeeds to identify the nonlinear effort in two model example cases up to the discretization error of the provided training data. First tests promise good generalizing properties outside of the provided test data.

It is worth noting that the proposed approach using Gaussian processes has no difficulty in identifying pH-DAE systems with a higher index - the constrained multibody system has index 3.

Future work needs to cover larger and more realistic test cases, where also larger parts of the pH-DAE, i.e.~not only the effort function, are identified. While the presented study could be easily executed on a laptop, future work also needs to cover fast methods for multi-task GPR, in order to reduce the high computational complexity for larger-scale applications.
\bibliographystyle{siam}

\bibliography{biblio.bib}
\end{document}

%% file: circuit.tex
\begin{tikzpicture}
	\begin{circuitikz}
		\draw[fill=black] (0,3) ellipse (.07 and .07) node[above,yshift=0.1cm]{$e_{1}$};
  \draw[fill=black] (4,3) ellipse (.07 and .07) node[above,yshift=0.1cm]{$e_{2}$};
		\draw (0,3) -- (1,3);
  \draw (3,3) -- (4,3);
		\draw (0,3) to[C, l=$q_C$] (0,0);
       \draw (4,3) to[voltage source, v=$u(t)$,i>^=$I_V$] (4,0);
  \draw (1,3) to[R, l=$G$] (3,3);
		\draw (0,0) -- (4,0);

		\begin{scope}[xshift=-1cm]
		\draw (3.,0) -- (3.,-0.6);
		\draw (2.6,-.6) -- (3.4,-.6);
		\draw (2.75,-.7) -- (3.25,-.7);
		\draw (2.9,-.8) -- (3.1,-.8);
		\end{scope}
		
	\end{circuitikz}
\end{tikzpicture}

%% file: pendulum.tex
\begin{tikzpicture}[angle radius=1.25cm]
	\begin{circuitikz}
		\draw[fill=black] (4,4) ellipse (.07 and .07);
\node (T) at (3.9,4.3) [left] {$(0,0)$};
  \draw (1,4) -- (7,4); 
  \draw (4,4) -- (6,1);
		\draw[fill=black] (6,1) ellipse (.4 and .4);
  \draw (4,4) -- (4,.2);
 \node (A) at (7,1)     [left]   {$m$};
 \coordinate (X) at (4,0);
  \coordinate (Y) at (4,4);
  \coordinate (Z) at (6,1);
  \path
  pic ["$\alpha$",  draw, fill=black!30] {angle = X--Y--Z};
 \draw[->]        (4,4)   -- (7,4);
   \node (AA) at (7.5,4)     [left]   {$x$};
    \draw[->]        (4,4)   -- (4,5);
   \node (AA) at (4.2,5.2)     [left]   {$y$};
   \draw[->]        (7,2.5)   -- (7,1.5);
   \node (AA) at (7.1,2.)     [right]   {$ m(\tilde g-u(t))$};
    \node (AAA) at (5.5,2.5)     [left]   {$l$};
	\end{circuitikz}
\end{tikzpicture}